\documentclass[lettersize,journal]{IEEEtran}

\usepackage{amsmath, amssymb, amsfonts}
\usepackage{algorithm, algcompatible}
\usepackage{url}
\usepackage{array}
\usepackage{graphicx}
\usepackage{cases}
\usepackage{caption}
\usepackage{subcaption}
\usepackage[english]{babel}

\usepackage{adjustbox}
\usepackage{textcomp}
\usepackage{verbatim}
\usepackage{multirow}

\usepackage{cite}
\usepackage{color, colortbl}
\usepackage{bm}     

\usepackage{hyperref}
\hypersetup{
	colorlinks=true,
	linkcolor=blue,
	filecolor=magenta,      
	urlcolor=cyan
}

\usepackage{acro}   

\makeatletter
\def\hlinewd#1{%
	\noalign{\ifnum0=`}\fi\hrule \@height #1 \futurelet
	\reserved@a\@xhline}
\makeatother

\newcolumntype{?}{!{\vrule width 1.5pt}}

\usepackage{pgfplots}
\usepackage{pgf}
\usepackage{tikz}
\usetikzlibrary{matrix}

\usetikzlibrary{fit}					  
\usetikzlibrary{backgrounds}	          
\usetikzlibrary{shapes.arrows}
\usetikzlibrary{shapes.geometric}
\usetikzlibrary{shapes.multipart}
\usetikzlibrary{matrix, arrows, positioning}
\tikzstyle{rec}=[draw,rectangle, minimum height=2cm]
\tikzset{>=stealth', punkt/.style={rectangle, 
		fill=gray!40, 
		draw=black, very thick, text width=3em, minimum height=2.5em, text centered}}
\tikzset{>=stealth', Denoi/.style={rectangle, fill=blue!20, 
		draw=black, very thick, text width=6em, minimum height=3.5em, text centered}}
\tikzset{>=stealth', CG/.style={rectangle, 
		fill=green!20, draw=black, very thick, text width=9.5em, minimum height=3.5em, text centered}}
\tikzstyle{background} = [rectangle, fill=green!20, inner sep=0.1cm, rounded corners=4mm, 
minimum height=6em]
\tikzstyle{sum}   = [draw, fill=gray!40, circle, node distance=1cm]
\tikzstyle{dot}   = [circle, fill=black, inner sep=0pt, minimum size=5pt, node contents={}]
\tikzstyle{fig_n} = [node distance=30pt, inner sep=0cm]

\def\*#1{\mathbf{#1}}
\def\+#1{\mathcal{#1}}
\def\-#1{\mathbb{#1}}
\def\~#1{\mathrm{#1}}
\def\R{\mathbb{R}}

\DeclareMathOperator{\diag}{diag}

\algnewcommand\INPUT{\item[\textbf{Input:}]}
\algnewcommand\OUTPUT{\item[\textbf{Output:}]}

\newcommand{\boldr}{\bm{r}}

\newcommand{\boldmu}{\bm{\mu}}

\newcommand{\boldA}{\bm{A}}

\newcommand{\calL}{\mathcal{L}}

\newcommand{\rmd}{\mathrm{d}}

\renewcommand{\hbar}{\bar{h}}

\newcommand{\argmin}{\operatornamewithlimits{arg\,min}}

\newcommand{\transp}{^\top}

\newcommand{\specialcell}[2][c]{%
	\begin{tabular}[#1]{@{}c@{}}#2\end{tabular}}

\DeclareAcronym{CT}{
	short=CT, 
	long=computed tomography,
}

\DeclareAcronym{DECT}{
	short=DECT, 
	long=ual-energy computed tomography,
}

\DeclareAcronym{FBP}{
	short=FBP, 
	long=filtered back projection,
}

\DeclareAcronym{MBIR}{
	short=MBIR, 
	long=model-based iterative reconstrution,
}

\DeclareAcronym{PWLS}{
	short=PWLS, 
	long=penalized weighted least squares,
}

\DeclareAcronym{PnP}{
	short=PnP, 
	long=Plug-and-Play Priors,
}

\DeclareAcronym{MAP}{
	short=MAP,
	long=maximum-a-posteriori,
}

\DeclareAcronym{CNN}{
	short=CNN, 
	long=convolutional neural network,
}

\DeclareAcronym{ART}{
	short=ART, 
	long=algebraic reconstruction techniques,
}

\DeclareAcronym{DL}{
	short=DL,
	long=deep learning,
}

\DeclareAcronym{MBDL}{
	short=MBDL,
	long=model-based DL,
}

\DeclareAcronym{CG}{
	short=CG,
	long=Conjugate Gradient,
}

\DeclareAcronym{DMDL}{
	short=DEcomp-MoD,
	long=Dual-Energy Decomposition Model-based Diffusion,
}

\DeclareAcronym{2D}{
	short=2-D,
	long=2-dimensional,
}

\DeclareAcronym{3D}{
	short=3-D,
	long=3-dimensional,
}

\DeclareAcronym{nD}{
	short=$n$-D,
	long=$n$-dimensional,
}

\DeclareAcronym{PET}{
	short=PET, 
	long=positron emission tomography,
}

\DeclareAcronym{MRI}{
	short=MRI, 
	long=magnetic resonance imaging,
}

\DeclareAcronym{MR}{
	short=MR, 
	long=magnetic resonance,
}

\DeclareAcronym{PCCT}{
	short=PCCT, 
	long=photon-counting computed tomography,
}

\DeclareAcronym{WLS}{
	short=WLS, 
	long=weighted least squares,
}

\DeclareAcronym{CS}{
	short=CS,
	long=compressed sensing,
}

\DeclareAcronym{TV}{
	short=TV,
	long=total variation,
}

\DeclareAcronym{TNV}{
	short=TNV, 
	long=total nuclear variation,
}

\DeclareAcronym{JTV}{
	short=JTV, 
	long=joint total variation,
}

\DeclareAcronym{DTV}{
	short=DTV, 
	long=directional total variation,
}

\DeclareAcronym{PLS}{
	short=PLS, 
	long=parallel level sets,
}

\DeclareAcronym{SQS}{
	short=SQS, 
	long=separable quadratic surrogate,
}

\DeclareAcronym{ADMM}{
	short=ADMM, 
	long=alternating direction method of multipliers,
}

\DeclareAcronym{CDL}{
	short=CDL,
	long=convolutional dictionary learning,
}

\DeclareAcronym{MCDL}{
	short=CDL,
	long=multichannel convolutional dictionary learning,
}

\DeclareAcronym{CAOL}{
	short=CAOL, 
	long=convolutional analysis operator learning,
}

\DeclareAcronym{MCAOL}{
	short=MCAOL, 
	long=multichannel convolutional analysis operator learning,
}

\DeclareAcronym{GAN}{
	short=GAN, 
	long=generative adversarial network,
}

\DeclareAcronym{VAE}{
	short=VAE, 
	long=variational autoencoder,
}

\DeclareAcronym{NLL}{
	short=NLL, 
	long=negative log-likelihood,
}

\DeclareAcronym{MSE}{
	short=MSE, 
	long=mean square error,
}

\DeclareAcronym{DDPM}{
	short=DDPM, 
	long=Denoising Diffusion Probabilistic Models,
}

\DeclareAcronym{DDIM}{
	short=DDIM, 
	long=Denoising Diffusion Implicit Models,
}

\begin{document}

\title{Direct Dual-Energy CT Material Decomposition using Model-based Denoising Diffusion Model}

\author{Hang Xu, Alexandre Bousse and Alessandro Perelli
\thanks{This work involved human subjects in its research. Approval of all ethical and experimental procedures and protocols for the AAPM low-dose CT dataset was granted by Mayo Clinic (USA) and it is made available through TCIA restricted license with de-identifiability of the data which is agreed by authors.}
\thanks{This work was supported by the Royal Academy of Engineering under the RAEng / Leverhulme Trust Research Fellowship LTRF-2324-20-160 and the French National Research Agency (ANR) under grant No ANR-20-CE45-0020.} 
\thanks{H. Xu was with the Department of Biomedical Engineering, University of Dundee, Scotland, DD1 4HN (UK) and he is currently with Longgang District Maternity \& Child Healthcare Hospital of Shenzhen City, Longgang Maternity and Child Institute of Shantou University Medical College (China).} 
\thanks{A.Perelli is with the Center for Medical Engineering and Technology, University of Dundee, Scotland, DD1 4HN (UK).}  
\thanks{A. Bousse is with University Brest, LaTIM, Inserm, U1101, 29238~Brest, France.}
\thanks{Corresponding author: A. Perelli, \mbox{\texttt{aperelli001@dundee.ac.uk}}}
} 

\markboth{}%
{Hang Xu \MakeLowercase{\textit{et al.}}: Direct Dual-Energy CT Material Decomposition using Model-based Denoising Diffusion Model}

\maketitle

\begin{abstract}
Dual-energy X-ray Computed Tomography (DECT) constitutes an advanced technology which enables automatic decomposition of materials in clinical images without manual segmentation using the dependency of the X-ray linear attenuation with energy. However, most methods perform material decomposition in the image domain as a post-processing step after reconstruction but this procedure does not account for the beam-hardening effect and it results in sub-optimal results. In this work, we propose a deep learning procedure called \ac{DMDL} for quantitative material decomposition which directly converts the DECT projection data into material images. The algorithm is based on incorporating the knowledge of the spectral DECT model into the deep learning training loss and combining a score-based denoising diffusion learned prior in the material image domain. Importantly the inference optimization loss takes as inputs directly the sinogram and converts to material images through a model-based conditional diffusion model which guarantees consistency of the results. We evaluate the performance with both quantitative and qualitative estimation of the proposed \ac{DMDL} method on synthetic DECT sinograms from the low-dose AAPM dataset. Finally, we show that \ac{DMDL} outperform state-of-the-art unsupervised score-based model and supervised deep learning networks, with the potential to be deployed for clinical diagnosis.
\end{abstract}


\section{Introduction}\label{sec:introduction}

\IEEEPARstart{D}{\ac{DECT}} is one spectral CT technology which is based on the deployment of two X-ray sources at different energies which can potentially allow to discriminate different materials in a specimen or to reconstruct virtual mono-energetic images which are of utmost interest in clinical imaging applications \cite{mendoncca2013flexible, liu2009quantitative}, industrial non-destructive testing \cite{sellerer2019quantitative} and airport explosive detection \cite{ying2006dual}. 

Different mechanical architectures can be deployed for \ac{DECT}: one option is to sequentially scan the specimen with the X-ray source operating either at low voltage (kVp) or high kVp. This approach is easy to implement from a hardware point of view but could suffer from misalignment due to the short delay between the contiguous scans \cite{marin2014state}. An alternative procedure relies on a fast kVp-switching technology which allows a single source to swap from high to low voltage in contiguous angular projections \cite{flohr2006first}. Finally, dual detector layers allows to acquire both low and high photon data simultaneously with correct spatial alignment \cite{vlassenbroek2011dual}.

The dependency of the attenuation coefficient of different materials with respect to the X-ray energy can be leveraged in the \ac{DECT} material decomposition procedure whose aim is to estimate each pixel's value as a linear combination of photoelectric absorption and Compton scattering \cite{alvarez1976energy} or using two different basis materials \cite{johnson2007material}. 

Different approaches have been developed to obtain material images: the image domain techniques are based on first reconstructing independently the energy-dependent attenuation in each pixel \cite{maass2009image}, then the high and low energy attenuation values can be approximated as a linear combination of three basis materials by imposing the constraint of volume conservation \cite{niu2014iterative}. Recently several deep learning networks have been deployed to fit the linear attenuation into basis material decomposition, such as the Butterfly-Net \cite{zhang2019image}, U-Net \cite{nadkarni2022material} and generative adversarial network (GAN) \cite{shi2021material} that are trained with a supervised learning approach which requires the pair of energy reconstructed \ac{DECT} images and basis material segmented images in the dataset. Although these methods are able to reduce the noise, they do not account for the beam-hardening effect, caused by the poly-energetic nature of the X-ray source and the propagation of errors from reconstruction to material decomposition \cite{bousse2023}. 

An alternative approach is based on decomposing the high and low-energy sinograms into two independent measurements which correspond to a single basis material. Different approximations of the decomposition function that convert the dual energy sinograms into materials independent sinograms have been proposed. A polynomial numerical approximation of the inverse of the poly-energetic measurement models was presented in \cite{alvarez1976energy}. Other works estimate the coefficients of the polynomial function to convert into materials' sinograms using empirical models \cite{cardinal1990accurate} or calibration phantoms \cite{zhang2013model}. Afterwards, each material sinogram is converted into the image domain using model-based optimization methods \cite{mechlem2018spectral} with spatial regularization. One challenge related to projection-based decomposition is that there is noise correlation between the decomposed material sinograms that should be estimated for the successive image reconstruction task. 

Several works have proposed an alternative one-step methodology which combines material decomposition and tomographic reconstruction to estimate material images directly from the energy sinograms. Although the one-step decomposition fully accounts for noise correlations in the \ac{DECT} model, it requires to solve a more difficult non-linear optimization problem in higher dimensional space which leads to a large increase in computational time. Numerous iterative solvers have been used such as approximate quadratic surrogate \cite{long2014,mechlem2018} and primal-dual algorithms \cite{foygelbarber2016}. 

Recently, other works have exploited the paradigm of combining deep learning and the knowledge of the physics of the \ac{DECT} model within the optimization problem. The one-step material decomposition is implemented using supervised unrolling algorithms in \cite{xia2023} or embedding the back-projection of the Radon transform in the network architecture \cite{su2022} and denoising the estimated material images at each iteration \cite{perelli2021}. 
Self-supervised approaches, which do not require the manually segmented material images as ground truth for training, have been developed for one-step decomposition \cite{fang2021} based on the Noise2Inverse framework which uses pairs of sub-sampled noisy sinograms and training dataset. 

Currently, diffusion models have shown state-of-the-art performance in image generation and in solving different inverse problems \cite{vazia2025}. Diffusion models for denoising task exploit the connection with score-based models and unconditional training as in \ac{DDPM} \cite{ho2020}. For inverse problems the generated images should be conditioned on the measurements which can be achieved either by training a new model given the measurements' condition explicitly \cite{saharia2022} or using the pre-trained unconditional diffusion model for conditional generation. However, the former method requires training new models for different tasks while the second method is more flexible. Denoising Diffusion Restoration Models (DDRM) \cite{kawar2022} proposes an unsupervised posterior sampling method but it needs to calculate the singular value decomposition (SVD) which is not computationally efficient in high-dimensional imaging problems. A method to solve sparse-view CT reconstruction is proposed in \cite{song2021} by constructing a tractable conditional sampling diffusion process and for material decomposition with \ac{DECT} in \cite{vazia2024spectral,jiang2024multi}. 

Recently plug-and-play approaches in \cite{Graikos2022} and (DiffPIR) \cite{Zhu2023} have been proposed which takes inspiration from exploiting any black-box denoisers as implicit image priors where the diffusion models can be used as generative denoisers. 

It is worth noting that most of the proposed diffusion model-based approaches for CT reconstruction have been applied to simple mono-energetic problems and we aim to develop a new model-based optimization framework combined with score-based diffusion prior to directly decompose materials' images from \ac{DECT} sinograms.

\subsection{Main Contribution}

In this work we propose a novel method for material decomposition from \ac{DECT} measurement using a projection-based decomposition and denoising diffusion model as image prior. The innovation is based on formulating the model-based material decomposition problem into decoupled sub-optimization problems operating on different data spaces:
\begin{itemize}
	\item the material decomposition is based in the measurements domain and the mapping from \ac{DECT} sinograms is learned through a model-based supervised learning approach. 
	\item the material images' prior is designed through a plug-and-play denoising approach based on a simple unconditional diffusion model.
\end{itemize}

This framework allows to learn separately two different networks with different input spaces and to guarantee the measurement consistency by external conditions on the diffusion denoising process during training. We show state-of-the-art results on synthetic low-dose \ac{DECT} sinograms obtained from real images from the Low-dose AAPM dataset \cite{mccollough2017}. 

\subsection{Organization of the Paper and Notation}

The paper is organized as follows: Section \ref{sec: DECT_model} introduces the forward model of DECT. Section \ref{sec: MBIR} describes the model-based optimization problem for DECT. Section \ref{sec:Diffusion} reviews the Denoising Diffusion Probabilistic Models (DDPM). Section \ref{sec:DEcomp-MoD} introduces our proposed DECT Material Decomposition using Model-Based Diffusion (\ac{DMDL}). Section \ref{sec:results} presents the results of proposed method. Section \ref{sec: discussion} discusses the problems in proposed method. 

Throughout the paper, we adopt the following notations: matrices or discrete operators and column vectors are written respectively in capital and normal boldface type, i.e. $\*A$ and $\*a$, to distinguish from scalars and continuous variables written in normal weight; $[\*a]_j$ denotes the $j$-th entry of $\*a$; `$(\cdot)\transp$' denotes the transposition and $\diag()$ a diagonal matrix.

\section{Dual-Energy CT Forward Model} \label{sec: DECT_model}

We introduce the forward mathematical model of the \ac{DECT} system where two sources emit poly-energetic X-ray photons with spectrum $S_k(E)$, where $k\in\{e_1,e_2\}$. The linear attenuation image takes the form of a spatially- and energy-dependent function $\mu(\boldr,E) \colon \R^2 \times \R_+ \to \R_+$ at position $\boldr\in\R^n$ and energy $E\in \R_+$. \Ac{DECT} system performs measurements along a collection of rays where $\calL_n \subset \R^n$ denotes the $n$-th ray, $n=1,\dots,N$ with $N$ being the number of detector pixels and $N_\rmd = 2$ the number of X-ray sources. We consider the \ac{DECT} system, where the angular position of the rays is the same for both acquisitions with different sources, therefore the geometry of the rays $\{\calL_n\}$ does not depend on $k$. 

The conditional mean of the photons' intensities $I_{n,k}$ given the linear attenuation $\mu(\cdot)$ can be modeled for all $n=1,\dots, N$ in the continuous domain using the Beer's law as following
\begin{equation}\label{eq:Beer_Law}
	\bar{I}_{n,k} = \-E\left[I_{n,k} | \mu\right] = \int_{0}^{\infty} S_k(E) \,\~e^{-\int_{\calL_n}\mu(\*r,E)dr} \,\~dE
\end{equation}
where `$\int_{\calL_j}$' denotes the line integral along $\calL_j$, $E$ (keV) is the photon energy, $S_k(E)$ is the overall spectrum for the source $k$, which consists by the normalized photon energy distribution for the source $k$ multiplied by the detector response and the linear attenuation $\bm\mu(\*r,E)$. 

The energy-dependent image is sampled on a grid of $M$ pixels, assuming that $\mu$ can be decomposed on a basis of $M$ ``pixel-functions'' 
$u_m$ such that
\begin{equation}\label{eq:decomp_disc}
	\mu(\boldr,E) = \sum_{m=1}^M \mu_m(E) u_m(\boldr) \, , \quad \forall (\boldr,E)\in\R^2\times\R^+ 
\end{equation} 
where $\mu_m(E)$ is the energy-dependent attenuation at pixel $m$. The line integrals in Eq.~\eqref{eq:Beer_Law} can be therefore rewritten as
\begin{equation}\label{eq:syst_mat}
	\int_{\calL_n} \mu(\boldr,E) \,\rmd \boldr  = [\boldA \boldmu(E)]_n 
\end{equation} 
with $\boldA \in \R^{N\times M}$ defined as $[\boldA]_{n,m} = \int_{\calL_n} u_m(\boldr) \,\rmd \boldr$ and $\boldmu(E) = [\mu_1(E),\dots,\mu_J(E)]\transp \in \R_+^M$ is the discretized energy-dependent attenuation. 
Substituting the Eq.~\eqref{eq:decomp_disc} in Eq.~\eqref{eq:Beer_Law}, we obtain the following discrete model 
\begin{equation}\label{eq:cond_count}
	\-E\left[I_{n,k} | \bm\mu\right] = \int_{0}^{\infty} S_k(E) \,\~e^{-\sum_{m=1}^M A_{n,m} \mu_m(E)} \,\~dE
\end{equation} 

In the normal dose case, the \ac{DECT} measurements collected from the scanner can be obtained using the negative logarithm of the measured photons at each energy as follows
\begin{equation}\label{eq:mean_photon}
	\*y_n = [y_{n,1}, y_{n,2}] = \left[  -\log\left( I_{n,1} \right), -\log\left( I_{n,2} \right)\right]
\end{equation}
where $\*y\in\-R^{N\times 2}$ and each row consists respectively of the low and high energy projections for the $n$-th ray, $\*y_n = \left[ y_{n,1}, y_{n, 2} \right]$. 

Using the formulation in \cite{thibault2006} and considering the first order Taylor expansion of the energy projection $\*y_n$, 
\begin{equation}
	y_{n,k} \approx -\log\left(\bar{I}_{n,k}\right) + \left(1 - \frac{I_{n,k}}{\bar{I}_{n,k}}\right)
\end{equation}
we approximate the conditional mean of the measurements over the linear attenuation $\-E\left[\*y_n | \bm\mu\right]$ as follow
\begin{equation}\label{eq:log_avg}
	\-E\left[\*y_n | \bm\mu\right] \approx -\log\left(\int_{0}^{\infty} \*S(E) \~e^{-\left[\*A\bm\mu(E) \right]_n} \~dE\right) \triangleq \hat{\*y}_n(\bm\mu)
\end{equation}
with $\*S(E) = \left[ S_{1}(E), S_{2}(E) \right]$ representing the spectrum of the two X-ray sources at energy $E = \{e_1, e_2\}$, and 
\begin{equation}
	\mathrm{Cov}(\*y_n | \bm\mu) = \left[
	\begin{array}{cc}
		\frac{1}{I_{n,1}} & 0 \\
		0 & \frac{1}{I_{n,2}}
	\end{array}
	\right]
\end{equation} 

From the model \eqref{eq:log_avg} connecting the measurement $\*y$ with attenuation $\bm\mu$, we can derive the approximated conditional mean $\hat{\*y}_n(\*x)$ 
over the vectorized material images $\*x\in\-R^{M\times 2}$, which contains the densities of the input object for the selected basis materials at the $m$-th voxel, $\*x_m = [x_{m,1}, x_{m,2}]$, by exploiting the decomposition of the linear attenuation as a linear combination of the mass density of two basis materials
\begin{equation}\label{eq:m_decomp}
	\mu_i(E) = x_{m,1}\varphi_1(E) + x_{m,2}\varphi_2(E)
\end{equation}
where $x_{m,s}$ (mg/cm$^3$) is the equivalent density, not dependent on the energy $E$, for basis materials $s=1,2$ at voxel $m$ and $\varphi_s(E)$ (cm$^2$/mg) is the known energy-dependent mass attenuation function for basis material $s$. 
Then, by substituting Eq.~\eqref{eq:m_decomp} into Eq.~\eqref{eq:log_avg}, we have
\begin{align}\label{eq:h} 
	& \-E \left[\*y_n | \*x\right] \nonumber \\
	& = -\log\left(\int_{0}^{\infty} \*S(E) \~e^{-\sum_{m=1}^M A_{n,m} \left(x_{m,1}\varphi_1(E) + x_{m,2} \varphi_2(E)\right)} \~dE\right) \nonumber \\
	& = -\log\left(\int_{0}^{\infty} \*S(E) \~e^{-\*p_n\left(\bm\varphi(E)\right)^T} \~dE\right) = h(\*p_n) 
\end{align}
where $\bm\varphi(E) = \left[\varphi_1(E), \varphi_2(E)\right]$, $\*p_n$ (mg/cm$^2$) is the material density projection, representing the line integral of material densities along ray $n$, defined as
\begin{equation}
	\*p_n = \left[\sum_{m=1}^M A_{n,m}x_{m,1}, \sum_{m=1}^M A_{n,m}x_{m,2}\right] = \left[\*A\*x\right]_n
\end{equation}
and the vector-valued function, $h: \-R^2 \rightarrow \-R^2$ models the non-linear relationship between the material density projections and the expected photon attenuation.  

The inverse function $h^{-1}: \-R^2 \rightarrow \-R^2$ is defined as
\begin{equation}\label{eq:h_inv}
    h^{-1}\left(h(\*p_n) \right) = \*p_n
\end{equation}
and it represents the material decomposition in the sinogram domain since it converts the energy sinogram $h(\*p_n)$ into the material sinograms $\*p_n$. In Sections \ref{sec: MBIR}-\ref{sec:Diffusion}, we will describe the individual terms exploited in the proposed \ac{DMDL} optimization problem derived in Section \ref{sec:DEcomp-MoD}, from the model representing the data fit term from energy sinogram to material images and the image prior based on diffusion models.

\section{Model-based Optimization Problem} \label{sec: MBIR}

In the normal X-ray dose case, we can approximate the Poisson distribution of the measurements $I_{n,k}$ with an anisotropic Gaussian distribution over $\*y_n$ with conditional average $\hat{\*y}_n(\*x)$ 
and diagonal covariance $\*W$, where individual projections are conditionally independent. The negative log-likelihood $f(\*y, \*x) = -\log P(\*y | \*x)$ is expressed as follow
\begin{equation}\label{eq:NLL}
	f(\*y, \*x) = \frac{1}{2} \sum_{n=1}^N\| \*y_n - \hat{\*y}_n(\*x) 
	\|^2_{\*W_n} + C
\end{equation}
where $C$ is a normalizing constant, and $\*W_n$ is the inverse covariance of $\*y_n$, $\*W_n = \~{Cov}^{-1}(\*y_n | \*x) = \diag(w_{n,1}, w_{n,2})$ with $w_{n,k} = \frac{1}{\~{Var}(\*y_{n,k} | \*x)} \approx I_{n,k}$, for $k \in \{ 1, 2 \}$. Substituting the expression of the conditional mean \eqref{eq:h} in \eqref{eq:NLL}, we obtain the explicit expression of the \ac{NLL} function
\begin{equation}\label{eq:NLL_h}
	f(\*y, \*x) = \frac{1}{2} \sum_{n=1}^N\| \*y_n - h\left([\*A\*x]_n\right) \|^2_{\*W_n} + C
\end{equation}

\Ac{MBIR} reconstruction approaches aim at optimising the \ac{MAP} cost function using \eqref{eq:NLL_h}, however this problem is difficult to solve for the following reasons: the function $f(\*y, \*x)$ is non-linear because of the mapping $h$ and direct optimisation is generally very computationally demanding and difficult since the objective does not have a quadratic form. Furthermore, the non-linear function $h$ is measured on the real \ac{DECT} scanner and the inverse function $h^{-1}$ is challenging to compute since it requires an accurate calibration procedure.

\section{Review Denoising Diffusion Probabilistic Models}\label{sec:Diffusion}

\begin{figure}[!h]
	\centering
	\includegraphics[width=.5\textwidth]{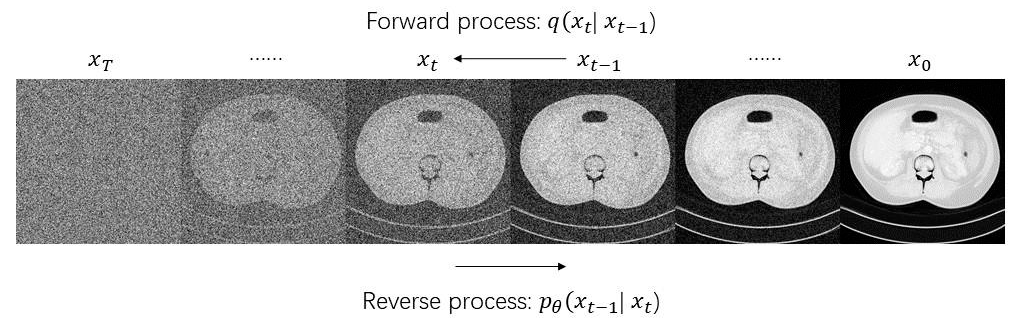}
	\caption{Schematic of the diffusion model. The forward process adds noise gradually while the reverse process generates the clean image from noise using a model with parameters $\bm\rho$.}\label{fig:diffusion_process}
\end{figure}

A diffusion model can be represented as a Markov chain which is composed by a forward process and a reverse process as shown in Fig. \ref{fig:diffusion_process}. The forward process gradually adds noise to obtain a completely random noisy image with Gaussian distribution. Then the reverse process recovers a clean image from the Gaussian distribution using a parameterized Markov chain trained by a score-based non-linear model \cite{ho2020}.

\noindent The forward process is represented as follows:
\begin{equation}
	q(\*x_{1:T}|\*x_0) = \prod_{t=1}^Tq(\*x_t|\*x_{t-1})
\end{equation}
where $T$ is the total time steps. In the forward process, we add the Gaussian noise with a fixed variance schedule $\{\beta_t \in (0, 1)\}^T_{t=1}$. The injection of the noise in each time step can be described by 
\begin{eqnarray}\label{eq:forw_diff}
	q(\*x_t|\*x_{t-1}) &=& \mathcal{N}\left(\*x_t; \sqrt{1-\beta_t} \*x_{t-1}, \beta_t\*I\right)\nonumber\\
	\*x_t &=& \sqrt{1-\beta_t} \*x_{t-1} + \sqrt{\beta_t} \bm\epsilon_{t-1} 
\end{eqnarray}
where $\bm\epsilon_{t-1}\sim\+N(0, \*I)$ is the noise at the $t-1$ step. Based on the property of Gaussian distribution and the reparameterization trick, Eq. (\ref{eq:forw_diff}) can be further used to directly obtain $\*x_t$ at arbitrary step $t$ from $\*x_0$ as
\begin{eqnarray}\label{eq:forward_step}
	q(\*x_t|\*x_0) &=& \mathcal{N}\left(\*x_t; \sqrt{\bar{\alpha}_t} \*x_0, (1 - \bar{\alpha}_t)\*I\right)\nonumber\\
	\*x_t &=& \sqrt{\bar{\alpha}_t} \*x_0 + \sqrt{1 - \bar{\alpha}_t} \bm\epsilon 
\end{eqnarray}
where $\alpha_t = 1 - \beta_t$ and $\bar{\alpha}_t = \prod_{t=1}^{T}\alpha_t$. 

The reverse process recovers the clean image from the noise with a parameterized Markov chain, represented by 
\begin{eqnarray}
	p_{\bm\rho} (\*x_{0:T}) & = & p(x_T)\prod_{t=1}^{T} p_{\rho} (\*x_{t-1}|\*x_t)\nonumber \\ 
	p_{\bm\rho} (\*x_{t-1}|\*x_t) &=& \+N\left(\*x_{t-1}; \bm\mu_{\bm\rho}(\*x_t, t), \bm\Sigma_{\bm\rho} (\*x_t, t) \right) 
\end{eqnarray}
where $p_{\bm\rho} (\*x_{t-1}|\*x_t)$ is the Gaussian distribution whose mean and variance are indicated respectively by $\bm\mu_{\bm\rho}(\*x_t, t)$ and $\bm\Sigma_{\bm\rho} (\*x_t, t)$ which take $\*x_t$ and $t$ as inputs and are needed to predict the mean and variance at $\*x_{t-1}$. Regarding the reverse process, one drawback of the sampling of \ac{DDPM} is that the inference from $\*x_T$ to $\*x_0$ is step by step and is very time consuming. However the \ac{DDIM} \cite{song2020} which was proposed to transform Markovian processes into non-Markovian processes, has improved the efficiency of sampling at least 10\texttimes{} faster compared to \ac{DDPM}. Each reverse step in \ac{DDIM} is given by
\begin{align}\label{eq:DDIM_update}
	\*x_{t-1} = & \underbrace{\sqrt{\bar{\alpha}_{t-1}} \left( \frac{\*x_t-\sqrt{1 - \bar{\alpha}_t}\bm\epsilon_{\bm\rho}(\*x_t,t)}{\sqrt{\alpha_t}} \right)}_{\mathrm{predicted} \; \*x_0} \\
	 & + \underbrace{\sqrt{1 - \bar{\alpha}_{t-1} - \sigma_{\eta_t}^2}\bm\epsilon_{\bm\rho}(\*x_t,t)}_{\mathrm{direction \; to}\; \*x_t}\;\; + \underbrace{\sigma_{\eta_t}\bm\epsilon}_{\mathrm{random \; noise}} \nonumber
\end{align}
where $\bm\epsilon_{\bm\rho}(\*x_t,t)$ is implemented as a denoising neural network to predict the noise between $\*x_t$ and $\*x_0$, $\epsilon\sim\+N(0, \*I)$ and $\sigma_{\eta_t}$ is a free parameter controlling the generative process. The parameters $\bm\rho$ are learned by minimizing the loss function
\begin{equation}\label{eq:diff_train}
	L(\bm\rho) = \-E_{\*x_0, \bm\epsilon, t} \left[\| \bm\epsilon_t - \bm\epsilon_{\bm\rho} (\sqrt{\bar{\alpha}_t}\*x_0 + \sqrt{1 - \bar{\alpha}_t}\bm\epsilon_t, t)\|^2\right]
\end{equation}
where the network $\bm\epsilon_{\bm\rho}$ takes $\*x_t$ at random time-step $t$ as input and the predicted noise is compared with the true noise $\bm\epsilon_t$. The minimization problem (\ref{eq:diff_train}) can be related to denoising score matching over multiple noise scales with index $t$ \cite{song2019}. 

We will exploit the intuition behind the DiffPIR framework \cite{Zhu2023} to develop the direct model-based optimization combined with diffusion prior \ac{DDIM} to estimate the material decomposition from \ac{DECT} sinograms.

\section{Model-based Diffusion model for \ac{DECT} Material Decomposition (\ac{DMDL})} \label{sec:DEcomp-MoD}

We propose a new model-based optimization framework called \ac{DMDL} to directly estimate material images $\*x\in\R^{M\times 2}_+$ from \ac{DECT} measurements $\*y\in\R^{N\times 2}$. The innovation is based on the idea to split the optimization into independent sub-problems in order to separate the data consistency term, by directly embedding a learned approximation of the material decomposition function $h^{-1}$ from (\ref{eq:h}) into the \ac{NLL} optimization function, and learn the material image prior by exploiting a score-based diffusion model. 

The material decomposition of the images $\*x\in\R^{M\times 2}$ can be formulated as an optimization problem where we aim to minimize the following cost function 
\begin{equation}\label{eq:gen_opt}
	\*x^* = \argmin_{\*x\in\R^{M\times 2}_+} f_{\bm\theta}(\*y,\*x) + \lambda \+R_{\bm\rho}(\*x)
\end{equation}
which combines the term $f_{\bm\theta}(\*y,\*x)$, parameterized over the learned parameters $\bm\theta$ to ensure the consistency of the estimation with the measurement based on the negative log-likelihood model in \eqref{eq:NLL}, and the regularization term $\+R_{\bm\rho}(\*x)$ which enforces a particular structure on the materials estimate and it is learned through a diffusion model using pre-trained materials' data. We constrained the optimisation model to non-negative material images and $\lambda$ is the regularization parameter. 

We solve (\ref{eq:gen_opt}) using the Half-Quadratic Splitting (HQS) algorithm to separate the denoising and data consistency tasks. By introducing the auxiliary variable $\*z$, (\ref{eq:gen_opt}) becomes 
\begin{equation}
		\*x^* = \argmin_{\*x\in\R^{M\times 2}_+} f_{\bm\theta}(\*y,\*x) + \lambda \+R_{\bm\rho}(\*z),\quad s.t. \;\; \*z = \*x
\end{equation}
with the associated Lagrangian to minimize 
\begin{equation}
	\+L_{\mu} = f_{\bm\theta}(\*y,\*x) + \lambda R_{\bm\rho}(\*z) + \frac{\mu}{2}\|\*z - \*x \|^2
\end{equation}

Therefore, the HQS algorithm allows the split between prior and measurements as follows

\begin{subnumcases} {\label{eq:HQS}}
\*z^k \quad = \argmin_{\*z\in\R^{M\times 2}_+} \frac{1}{2(\sqrt{\lambda/\mu})^2} \left\| \*z - \*x^k \right\|^2 + \+R_{\bm\rho}(\*z) \label{eq:HQS_1}\\
\*x^{k-1} = \argmin_{\*x\in\R^{M\times 2}_+} f_{\bm\theta}(\*y,\*x) + \mu \left\| \*x - \*z^k\right\|^2 \label{eq:HQS_2}
\end{subnumcases}

It is worth noting that by formulating the problem with the split in \eqref{eq:HQS_1}-\eqref{eq:HQS_2}, we are able to decouple the learning procedures which are applied in different domains since the data consistent term contains the learning module $\+P_{\bm\theta}(\cdot)$ for the material decomposition in the sinogram domain \eqref{eq:HQS_2} while the image prior $\+R_{\bm\rho}(\cdot)$ is learned through the diffusion model computed in \eqref{eq:HQS_1}. 

In the following subsection we will detail the analytic formulation and how to solve each sub-problem.

\subsection{Model-based Material Decomposition Sub-problem}\label{sec:MBIR_decomp}

We want to develop a solver for the sub-problem \eqref{eq:HQS_2} directly from \ac{DECT} sinogram to material images. Therefore, considering the formulation in terms of attenuation images described in the \ac{NLL} function \eqref{eq:NLL_h}, we need to apply the inverse decomposition $h^{-1}$ in the projection domain as in Eq. \eqref{eq:h_inv}. The intuition is that the optimization sub-problem becomes linear and the function $h^{-1}$ is learned from the projection data during the iteration procedure of the solver.

Furthermore, the proposed method does not require any system calibration procedures to determine the material decomposition function $h^{-1}$ as needed in previous approaches. The designed cost function contains the \ac{NLL} term which ensures the consistency in the projection domain, achieved by mapping into a linear problem through the learned $h^{-1}$. 

\begin{figure}[!h]
	\centering
	\includegraphics[width=.5\textwidth]{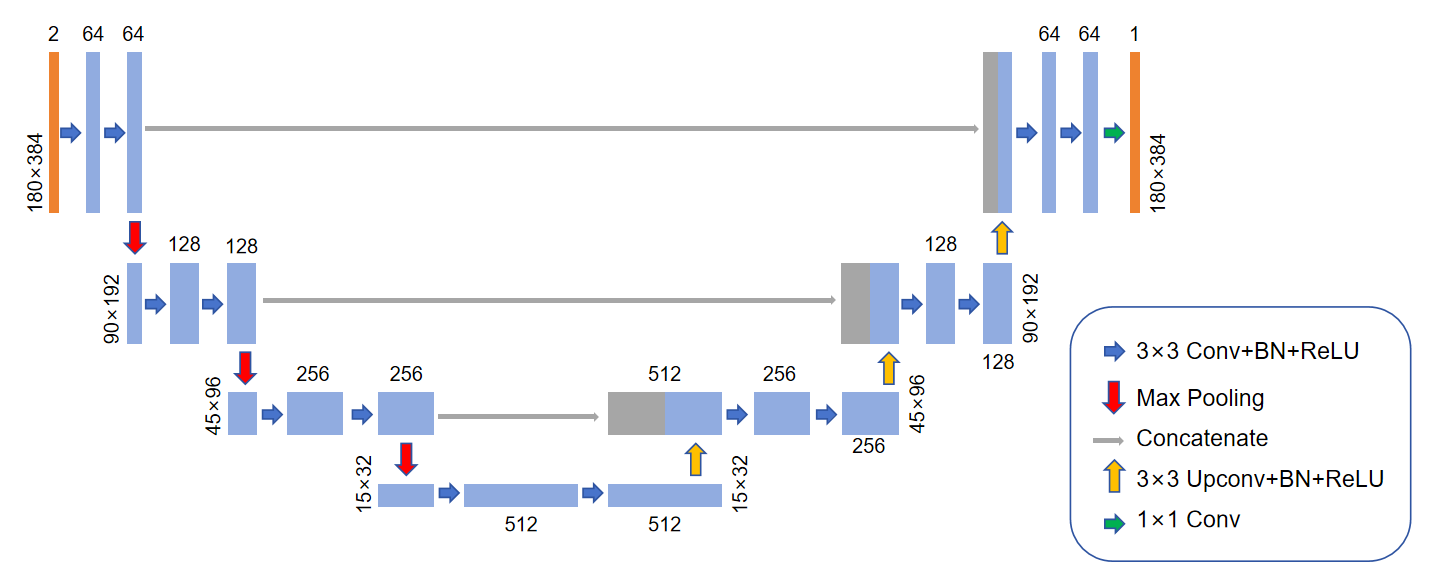}
	\caption{Schematic of the U-Net which implements the material decomposition function $\bm{\+P}_{\theta}$ in the projection domain.}\label{fig:U-net}
\end{figure}

We first describe the idea of embedding a learned material decomposition function $\+P_{\theta}$, where $\+P$ is a U-Net whose parameters $\theta$ are learned from the sinogram data, into the \ac{NLL} term. Using the definition \eqref{eq:h_inv}, we have
\begin{equation}
    h^{-1}\left([\*A\*x]_n \right) = [\*A\*x]_n
\end{equation}
and given the \ac{DECT} model \eqref{eq:h} and \eqref{eq:h_inv}, we obtain the estimated material sinogram as follow
\begin{equation}\label{eq:proj_p_decomp}
    \hat{\*p}_n = h^{-1}(\*y)_n \approx \bm{\+P}_{\theta}(\*y)_n 
\end{equation}
where the last approximation is obtained by substituting the vector function $\bm{\+P}_{\bm\theta} \colon \R^{N\times 2}_+ \to \R^{N\times 2}_+$ with learned parameters $\bm\theta$ for the non-linear material decomposition function $h^{-1}$. $\bm{\+P}_{\bm\theta}$ is implemented using a U-Net architecture which takes as input the energy sinogram $\*y\in\R^{N\times 2}$ and output the material sinogram of the same dimension as shown in Fig. \ref{fig:U-net}. The implementation's details of the network are described in Section \ref{sec:mat_Unet}.

We train the U-Net decomposition network in a supervised way where the output is compared with the respective material sinogram label. The parameters $\bm\theta$ are obtained by minimizing the following loss function 
\begin{equation}
	\hat{\bm\theta} = \argmin_{\bm\theta}\frac{1}{2}\| \bm{\+P}_{\bm\theta}(\*y) - \*p^*  \|^2
\end{equation}
where $\*p^*$ is the material sinogram label obtained by $\*p^* = \*A\*x^*$, with $\*x^*$ the ground truth material image. 

At inference, substituting \eqref{eq:proj_p_decomp} into \eqref{eq:NLL_h}, the \ac{NLL} function, which ensure consistency between \ac{DECT} sinograms $\*y$ and material images $\*x$, is given by 
\begin{equation}
    f_{\bm\theta}(\*y, \*x) = \frac{1}{2}\sum_{n=1}^{N} \|\left[\bm{\+P}_{\bm\theta}(\*y)\right]_n - [\*A \*x]_n \|^2_{\*B_n}
\end{equation}
where $\*B_n\in \R^{2\times 2}$ is the new statistical covariance matrix of the problem in the new domain which is obtained using the first order Taylor expansion as proposed in \cite{zhang2013model}. Using the polynomial network $\bm{\+P}_{\theta}$, we obtain the following expression for the covariance matrix 
\begin{equation}
    \*B_n = [\nabla \bm{\+P}_{\bm\theta}(\*y)_n]^{-1}\, \*W_n\, [\nabla \bm{\+P}_{\bm\theta}(\*y)_n]^{-T}
\end{equation}
where $\nabla \bm{\+P}_{\bm\theta}(\cdot)$ can be efficiently computed using back-propagation. The optimization sub-problem \eqref{eq:HQS_1} becomes
\begin{equation}\label{eq:opt_DECT}
    \*x^{k-1} = \argmin_{\*x\in\R^{M\times 2}_+} \frac{1}{2}\sum_{n=1}^{N} \left\|\bm{\+P}_{\bm\theta}(\*y)_n - \left[\*A \*x\right]_n \right\|^2_{\*B_n} + \mu \left\|\*x - \*z^k\right\|^2
\end{equation}

The optimization problem \eqref{eq:opt_DECT} has a quadratic form, therefore, the closed form solution is
\begin{equation}\label{eq:data_consist}
    \*x^k = \left[\*A^T(\*B\odot \*A) +\mu\*I\right]^{-1}\left( \*A^T \*B \, \bm{\+P}_{\bm\theta}(\*y) + \mu \*z^k \right)
\end{equation}
where $\*B=\left[ \diag(\*B_1) ; \ldots ; \diag(\*B_N)\right]\in\R^{N\times 2}$ is the concatenation of the diagonal element of the covariance weighting matrix for each sinogram pixel $n=1, \ldots, N$, $\odot$ represents the element-wise multiplication operation and $k = 1, \ldots, K$ is the iteration index. Although the close for solution \eqref{eq:data_consist} requires the computation of the inverse of the high dimensional matrix $\*G = \left[\*A^T(\*B\odot \*A) +\lambda\*I\right]\in\R^{M\times M}$, we can iteratively solve the linear problem 
\begin{equation}
    \*G\*x^k = \*A^T \*B \,\bm{\+P}_{\bm\theta}(\*y) + \mu \*z^k
\end{equation}
using the \ac{CG} algorithm \cite{nazareth2009conjugate} which does not require any computation of the inverse of $\*G$. 

\begin{figure}[!h]
	\centering
	\includegraphics[width=.5\textwidth]{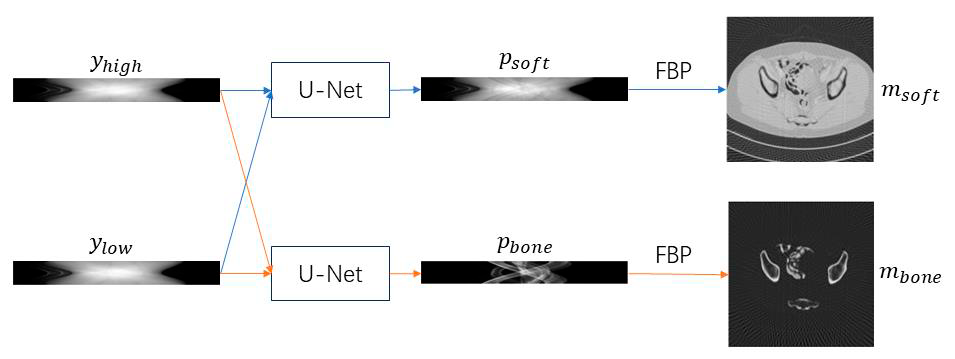}
	\caption{Diagram for the supervised training of the projection-based decomposition term $\*x_c = \*A^T \*B \, \bm{\+P}_{\bm\theta}(\*y)$ for two materials, soft tissue and bone, using the U-Net architecture.}\label{fig:U-Net_workflow}
\end{figure}

It is worth noting that the term dependent on the sinogram decomposition $\*x_c = \*A^T \*B \,  \bm{\+P}_{\bm\theta}(\*y)$ is not changing with the iteration index $k$ and it can be computed outside the iteration loop, since the parameters $\bm\theta$ are fixed at inference and computed during the training. The workflow of the projection-based material decomposition sub-problem is shown in Fig.~\ref{fig:U-Net_workflow} where $\*A^T$ is implemented using \ac{FBP}.

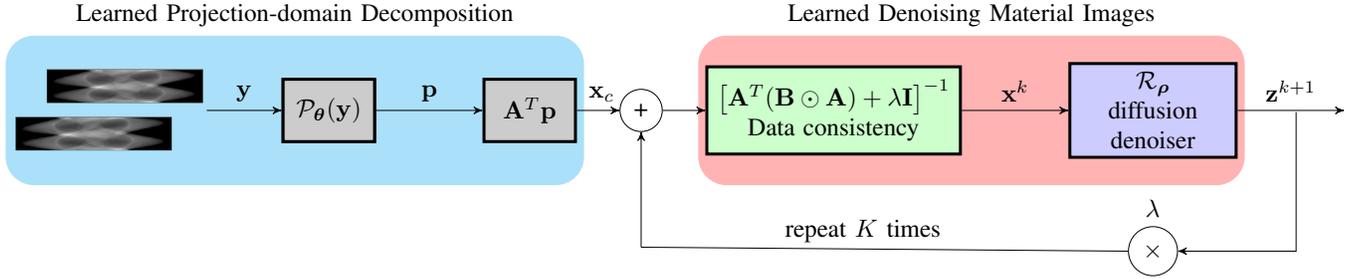
\begin{figure*}[!t]
	\centering
	\resizebox{\textwidth}{!}{
		\begin{tikzpicture}[node distance=1.5cm,scale=1]
			\node [punkt] (P) {$\+P_{\bm\theta}$};
			\node[fig_n, left= of P, inner sep=0pt] (IN) 
			{\includegraphics[width=.15\textwidth]{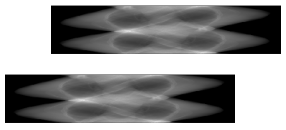}};	
			\path[->] (IN.east) edge node [above] {$\*y$} (P.west);
			\node [punkt, right= of P] (BP) {$\*A^T\*p$};
			\path[->] (P.east) edge node [above] {$\*p$} (BP.west);
			\node [punkt] (P) {$\+P_{\bm\theta}(\*y)$};
			\node[draw, fill=white, circle, right= 0.6cm of BP] (sum) {+};
			\node [CG, right= 0.6cm of sum] (DC) {$\left[\*A^T(\*B\odot\*A) +\lambda\*I\right]^{-1}$\\ Data consistency};	
			\draw[->] (sum.east) -- (DC.west);
			\path[->] (BP.east) edge node [above] {$\;\*x_c$} (sum.west);
			\node [Denoi, right= of DC] (Denoi) {$\+R_{\bm\rho}$\\ diffusion denoiser};
			\path[->] (DC.east) edge node [above] {$\*x^k$} (Denoi.west);
			\node [right= of Denoi, text width=0em, text centered] (OUT) {};
			\path[->] (Denoi.east) edge node [above] {$\*z^{k+1}$} (OUT.west);
			\node[label={$\lambda$}, draw, fill=white, circle, below= .95cm of Denoi] (prod) {$\times$};
			\node [below= of sum, text width=0em, text centered] (EMPTY)  {};
			\draw [->] (OUT.south)+(-0.8,0.1) |- (prod.east) [above] node{$\hspace{-9cm}$ repeat $K$ times};
			\path[-] (prod.west) edge node {} (EMPTY.center);
			\path[->] (EMPTY.center) edge node {} (sum.south);
			\begin{pgfonlayer}{background}[node distance=1cm]
				\node [background,fit=(IN) (P) (BP), fill=cyan!30, label=above:Learned Projection-domain Decomposition] (p_decomp) {};
			\end{pgfonlayer}
			\begin{pgfonlayer}{background}[node distance=1cm]
				\node [background,fit=(DC) (Denoi), fill=red!30, label=above:Learned Denoising Material Images] (m_denoi) {};
			\end{pgfonlayer} 
	\end{tikzpicture}}
	\caption{\ac{DMDL} algorithmic workflow for material decomposition. The energy sinogram $\*y$ is decomposed through $\+P_{\bm\theta}$ into material sinogram $\*p$ and back-projected to obtain the material images initial estimate $\*x$. The material images are iteratively denoised while being consistent with the measurements by using the denoising diffusion model  $\+R_{\bm\rho}$.}\label{fig:DMDL_workflow}
\end{figure*}

\subsection{Denoising Diffusion model Sub-Problem}

We solve the material image prior sub-problem \eqref{eq:HQS_1} by exploiting the denoising diffusion model in a plug-and-play approach. The objective is to solve the material decomposition problem using posterior sampling with diffusion model decoupled from the data fit term which constrains the material images to the measurements $\*y$. 

The sub-problem \eqref{eq:HQS_1} can be seen as a deep prior denoiser where we want to solve $\*z$ from the noisy vector $\*x^k$ given the noise level $\bar{\sigma}^k=\sqrt{\frac{1-\bar{\alpha}^t}{\bar{\alpha}^t}}$. Since \eqref{eq:HQS_1} can be solved as a proximal operator, this allows to draw a connection between the score model defined as 
\begin{equation}
	\nabla_{\*x} \+P(\*x) = - \nabla_{\*x} \log p (\*x) = - \*s_{\bm\rho}(\*x)
\end{equation}
and the \ac{DDIM} reverse step. Indeed if we consider using gradient descent, we can write
\begin{equation}
	\*z^t \approx \*x^t + \sqrt{\frac{1-\bar{\alpha}^t}{\bar{\alpha}^t}}\*s_{\bm\rho}(\*x^t)
\end{equation}
which correspond to the first term of the reverse step update in \eqref{eq:DDIM_update} where the score model $\*s_{\bm\rho}(\*x^t) = -\frac{\bm\epsilon_{\bm\rho}(\*x^t,t)}{\sqrt{1-\bar{\alpha}^t}}$ is connected to the pre-trained diffusion model $\bm\epsilon_{\bm\rho}(\*x^t,t)$. The predicted estimate $\*z^t$ at step $t$ is then conditioned on the \ac{DECT} measurements by solving the decomposition sub-problem using the CG algorithm as described in Section \ref{sec:MBIR_decomp}. 

To connect \eqref{eq:HQS} to the specific diffusion process at timestep $t = T, \ldots, 1$, we first define the diffusion noise level $\bar{\sigma}_t=\sqrt{\lambda/\mu}$ in \eqref{eq:HQS_1} and then rewrite \eqref{eq:HQS} as

\begin{subnumcases} {\label{eq:decomp-mod}}
\*z_0^t \; = \frac{1}{\sqrt{\bar{\alpha}_t}}\left(\*x_t + (1-\bar{\alpha}_t)\*s_{\bm\rho}(\*x_t, t)\right) \label{eq:decomp-mod1}
\\
\hat{\*x}_0^t \; = \argmin_{\*x\in\R^{M\times 2}_+} \sum_{n=1}^{N} \left\|\hat{\*p}_n - \left[\*A \*x\right]_n \right\|^2_{\*B_n} + \mu_t \left\|\*x - \*z^k\right\|^2 \label{eq:decomp-mod2}
\\
\*x_{t-1} = \sqrt{\bar{\alpha}_{t-1}}\hat{\*x}_0^t + \sqrt{1 - \bar{\alpha}_{t-1}}\left( \sqrt{1 - \xi} \hat{\bm\epsilon} +\sqrt{\xi}\bm\epsilon\right) \label{eq:decomp-mod3}
\end{subnumcases}
where $\mu_t = \lambda/\bar{\sigma}_t^2$, $\*z_0^t$ is the predicted clean image from diffusion model at time-step $t$ and $\hat{\*x}_0^t$ is the corrected image based on the material sinogram obtained from decomposition U-Net. One step of the update finishes with \eqref{eq:decomp-mod3} which is the modified version of \eqref{eq:DDIM_update}.

As described in the \ac{DDIM} reverse step in Eq. \eqref{eq:DDIM_update} the additional terms refer to adding the predicted noise obtained from Eq. \eqref{eq:forward_step}
\begin{equation}
	\hat{\bm\epsilon} = \frac{1}{\sqrt{1 - \bar{\alpha}_t}}\left(\*x_t - \sqrt{\bar{\alpha}_t}\hat{\*x}_0^t\right)
\end{equation}
and the random noise $\epsilon\sim\+N(0,\*I)$. A new parameter $\xi$ is introduced to control the trade off between predicted noise $\hat{\bm\epsilon}$ and the random noise $\bm\epsilon$, leading to the following update
\begin{equation}
	\*x_{t-1} = \sqrt{\bar{\alpha}_{t-1}}\hat{\*x}_0^t + \sqrt{1 - \bar{\alpha}_{t-1}}\left( \sqrt{1 - \xi} \hat{\bm\epsilon} +\sqrt{\xi}\bm\epsilon\right)         
\end{equation}
The DEcomp-MoD framework is summarised in Algorithm \ref{table:DMDL} and the schematic workflow of the training and inference procedure is shown in Fig. \ref{fig:DMDL_workflow}.

\begin{algorithm}[!h]
	\caption{\ac{DMDL} algorithm}
	\label{table:DMDL}
	\begin{algorithmic}[1]
		\REQUIRE $\*y$, pre-trained $\+P_{\bm\theta}$ and $\bm\epsilon_{\bm\rho}$, $\*A$, $T$, $\{\bar{\sigma}_t\}$, $\lambda$, $\xi$
		\INPUT  Initialize $\*x_T\sim\+N(0,\*I)$, $\mu_t = \lambda/\bar{\sigma}_t^2$, \\ 
		material sinogram decomposition $\hat{\*p} = \+P(\*y)$ \\
		pre-compute initial material estimate $\*x_c = \*A^T\*p$
		\OUTPUT $\*x_0$
		\FOR{$t = T, \ldots, 1$}
		\STATE $\*z_0^t = \frac{1}{\sqrt{\bar{\alpha}_t}}\left(\*x_t + (1-\bar{\alpha}_t)\*s_{\bm\rho}(\*x_t, t)\right)$
		\STATE $\hat{\*x}_0^t = \argmin_{\*x\in\R^{M\times 2}_+} \sum_{n=1}^{N} \|\hat{\*p}_n - [\*A \*x]_n \|^2_{\*B_n} + \mu_t \|\*x - \*z^k\|^2$,  solved using CG algorithm
		\STATE $\hat{\bm\epsilon} = \frac{1}{\sqrt{1 - \bar{\alpha}_t}}\left(\*x_t - \sqrt{\bar{\alpha}_t}\hat{\*x}_0^t\right)$
		\STATE $\epsilon\sim\+N(0,\*I)$
		\STATE $\*x_{t-1} = \sqrt{\bar{\alpha}_{t-1}}\hat{\*x}_0^t + \sqrt{1 - \bar{\alpha}_{t-1}}\left( \sqrt{1 - \xi} \hat{\bm\epsilon} +\sqrt{\xi}\bm\epsilon\right)$
		\ENDFOR
	\end{algorithmic}
\end{algorithm}

\section{Results}\label{sec:results}

\subsection{DECT Data Simulation}

\subsubsection{Data Source}

Given the scarcity of  clinical DECT raw measurement data, 
we simulated DECT energy sinograms and ground truth material density images from full-dose mono-energetic CT images. The datasets are obtained from the 2016 NIH-AAPM-Mayo Clinic Low-Dose CT Grand Challenge \cite{mccollough2017}. We used 5410 slices of CT images from 9 patients as training datasets and 526 slices from 1 patient as testing datasets. The image resolution was downsampled to 256\texttimes{}256.

\subsubsection{Simulation Methods}\label{sec: data_simulation}

The simulation of DECT energy-dependent sinograms is based on Eq. (\ref{eq:h}). The first step is to segment basis material images. We used the thresholding method \cite{kachelrieb2005} to segment bone- and water-basis material images, represented by $x_{m,1}$ and $x_{m,2}$. The material density sinograms were obtained using a forward projecting model $A_{n,m}$ which was simulated by Tomosipo \cite{hendriksen-2021-tomos}. We used a 2D fan-beam geometry where the number of detectors was set to 384 and the detector width was set to $1.5$ mm. The source-origin distance was set to $1000$ mm and the source-detector distance was set to $1500$ mm. We simulated the DECT acquisition system with fast low- and high-kVp switching between adjacent angles. Projections were collected from one rotation of $360$ degrees with interval of 1 degree, therefore each low- and high-energy transmission was sampled by $180$ angles. The incident photons were set to $2\cdot{} 10^6$ We aimed to use the generative ability of diffusion model to eliminate the artifacts caused by sparse-view sampling.

\begin{figure}[!h]
	\begin{center}	
		\includegraphics[width=.5\textwidth]{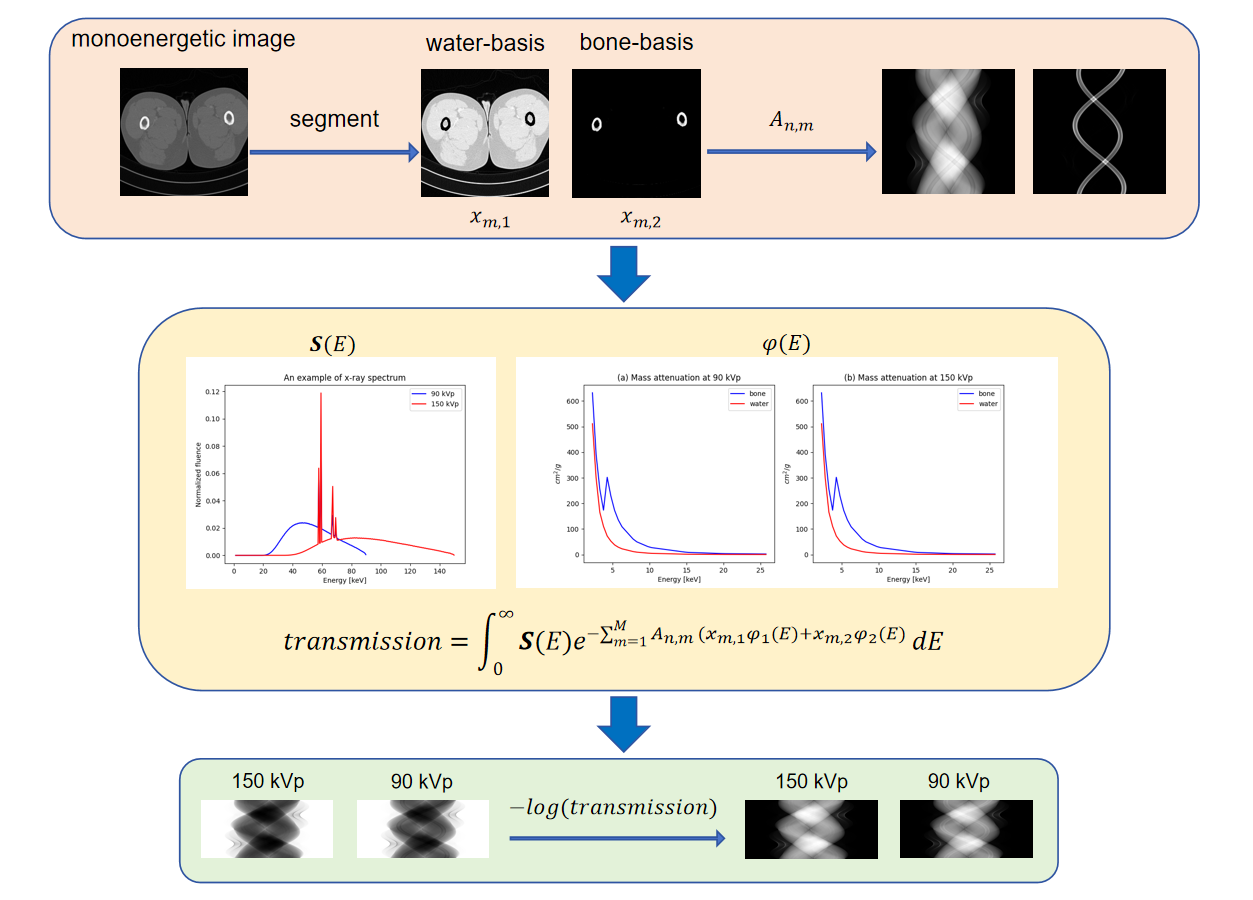}
		\caption{The schematic of DECT energy sinograms simulation. First, the mono-energetic CT image is segmented by selected thresholds to obtain the material density images $x_{m,1}$ and $x_{m,2}$. Then, the transmission CT measurements are simulated with the Beer's Law using the known projector $A_{n,m}$, normalized spectra $\*S(E)$ for $90$ kVp and $150$ kVp, and the mass attenuation coefficients $\varphi(E)$. Finally, the negative log of the transmission CT data is computed to obtain the energy-dependent sinograms at $90$ kVp and $150$ kVp.}
		\label{fig:data_simulation}
	\end{center}
\end{figure}

We used a Python package Spekpy \cite{poludniowski2021} to generate the two spectra $\*S(E)$ at 90 kVp and 150 kVp, respectively. The anode angle was set to 15 degrees. The spectrum at 90 kVp was generated with $1.5$ mm aluminum (Al) and $0.2$ mm copper (Cu) filtration. The 150 kVp spectrum was generated with 1.5 mm Al and 1.2 mm Cu filtration. The mass attenuation coefficients $\bm\varphi(E)$ of bone and water were obtained from the National Institute of Standards and Technology (NIST) report \cite{hubbell1995}. Substituted above values into Eq. (\ref{eq:h}) and we obtained low- and high-energy sinograms $\*y_n$. The overall process of DECT data simulation is shown in Fig. \ref{fig:data_simulation}.

\subsection{Training details}

\subsubsection{Material Decomposition}\label{sec:mat_Unet}

For the material decomposition problem, we used a U-Net architecture similar to \cite{chen2021} but without residual output. The inputs are two energy sinograms concatenated as one input with two channels and the output is a one-channel material sinogram. The details of the architecture are shown in Fig. \ref{fig:U-net}. 

The U-Net was trained on one NVIDIA RTX4090 GPU, with Pytorch \cite{paszke2019}, using 200 epochs and the batch size was 16. Adam optimizer with default settings was used except the start learning rate was set to $10^{-4}$ and it decayed by a factor of $0.98$ every $500$ steps. The mean square error (MSE) was used to compute the training loss.

\subsubsection{Diffusion Model}

We used the same network architecture as in DDPM \cite{ho2020}. For the model's settings we had $6$ feature map resolutions $(256, 256, 128, 64, 16, 4)$, $2$ convolutional residual blocks at each resolution and self-attention blocks at resolution 16\texttimes{}16.

The diffusion model was trained on one NVIDIA RTX4090 GPU, using the Pytorch framework. For the diffusion process, the variance schedule $\beta$ was set linearly from $0.0001$ to $0.02$, and the total time-steps $T$ was set to $1000$. Adam optimizer was used and the learning rate was set to $2\cdot{} 10^{-5}$. The models of the two materials were both trained for $6\cdot{} 10^5$ steps.

\subsection{Evaluation Settings and Metrics}

All the settings for the CT scanner and diffusion process were fixed as described in the previous sections. In the sampling settings, the number of iterations for CG was fixed to $10$. The parameters $\lambda$ and $\xi$ were set respectively to $0.001$ and $1.0$ as the default baseline. The effect of parameters will be discussed in section \ref{sec: free_parameters}. The sampling iterations of DEcomp-MoD were set to $100$ as default. Different sampling iterations and the running time will be evaluated in section \ref{sec: free_iters}. In our simulation, we used the peak-signal-to-noise ratio (PSNR) and the structural similarity index measure (SSIM) as quantitative metrics.

\begin{figure*}[!h]
	\begin{center}
		\small\addtolength{\tabcolsep}{-18pt}
		\renewcommand{\arraystretch}{0.1}
		\resizebox {.85\textwidth} {.85\height} {
			\begin{tabular}{ccccc}
				\hspace{-.6cm}\small (a) FBP 
				& 
				\hspace{-.6cm}\small (b) MCG diffusion
				& 
				\hspace{-.6cm}\small (c) DIRECT-Net 
				& 
				\hspace{-.6cm}\small (d) DEcomp-MoD
				& 
				\hspace{-.6cm}\small (e) Ground truth 
				\\
				\vspace{-.15cm} 
				\begin{tikzpicture}  
					\node {\includegraphics[width=0.2\textwidth]{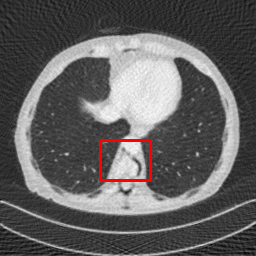}};
				\end{tikzpicture} \hspace{.55cm}
				&
				\begin{tikzpicture}
					\node {\includegraphics[width=0.2\textwidth]{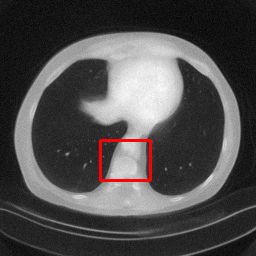}};
				\end{tikzpicture} \hspace{.55cm}
				&
				\begin{tikzpicture}
					\node {\includegraphics[width=0.2\textwidth]{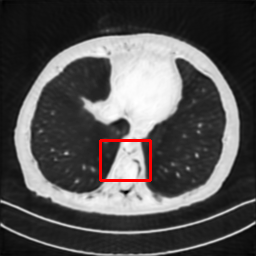}};
				\end{tikzpicture} \hspace{.55cm}
				&       
				\begin{tikzpicture}  
					\node {\includegraphics[width=0.2\textwidth]{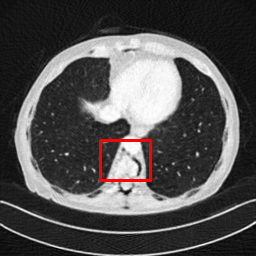}};
				\end{tikzpicture} \hspace{.55cm}
				&       
				\begin{tikzpicture}
					\node {\includegraphics[width=0.2\textwidth]{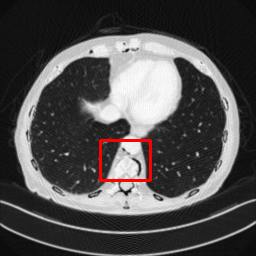}};
				\end{tikzpicture} \hspace{.55cm}
				\\   	
				\hspace{.03cm} 
				\begin{tikzpicture}  
					\node {\includegraphics[width=0.2\textwidth]{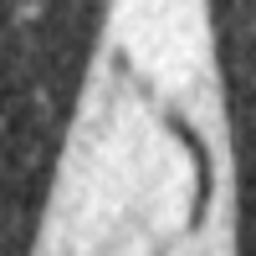}};
				\end{tikzpicture} \hspace{.55cm}
				&
				\begin{tikzpicture}
					\node {\includegraphics[ width=0.2\textwidth]{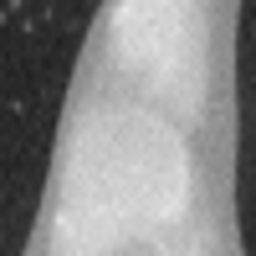}};
				\end{tikzpicture} \hspace{.55cm}
				&
				\begin{tikzpicture}
					\node {\includegraphics[width=0.2\textwidth]{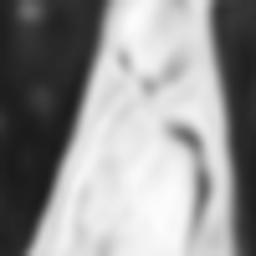}};
				\end{tikzpicture} \hspace{.55cm}
				&       
				\begin{tikzpicture}  
					\node {\includegraphics[width=0.2\textwidth]{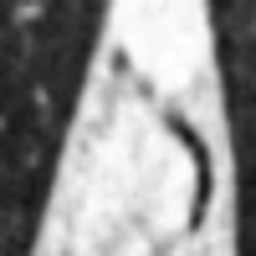}};
				\end{tikzpicture} \hspace{.55cm}
				&       
				\begin{tikzpicture}
					\node {\includegraphics[width=0.2\textwidth]{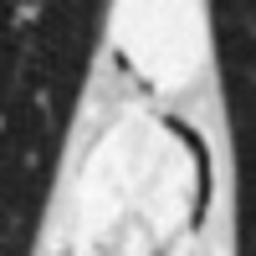}};
				\end{tikzpicture} \hspace{.55cm}
				\\
				\vspace{-.15cm} 
				\begin{tikzpicture}  
					\node {\includegraphics[width=0.2\textwidth]{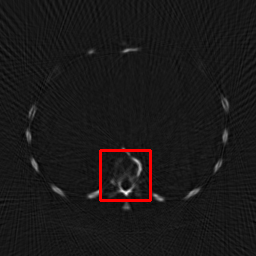}};
				\end{tikzpicture} \hspace{.55cm}
				&
				\begin{tikzpicture}
					\node {\includegraphics[width=0.2\textwidth]{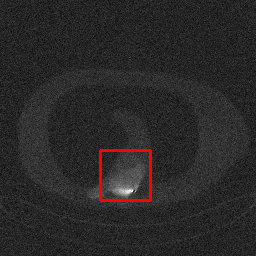}};
				\end{tikzpicture} \hspace{.55cm}
				&
				\begin{tikzpicture}
					\node {\includegraphics[width=0.2\textwidth]{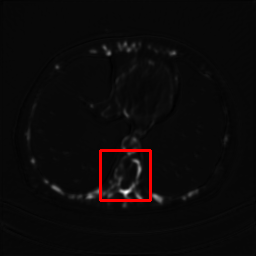}};
				\end{tikzpicture} \hspace{.55cm}
				&       
				\begin{tikzpicture}  
					\node {\includegraphics[width=0.2\textwidth]{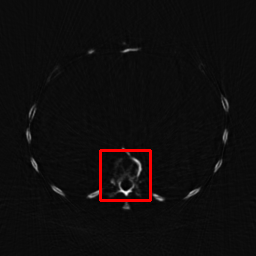}};
				\end{tikzpicture} \hspace{.55cm}
				&       
				\begin{tikzpicture}
					\node {\includegraphics[width=0.2\textwidth]{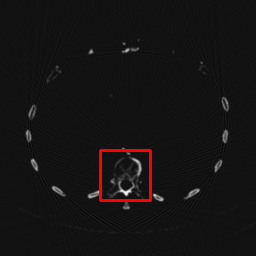}};
				\end{tikzpicture} \hspace{.55cm}
				\\   	
				\hspace{.03cm} 
				\begin{tikzpicture}  
					\node {\includegraphics[width=0.2\textwidth]{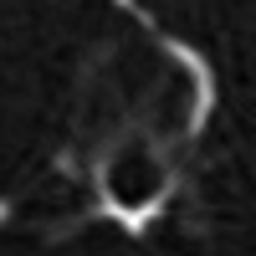}};
				\end{tikzpicture} \hspace{.55cm}
				&
				\begin{tikzpicture}
					\node {\includegraphics[ width=0.2\textwidth]{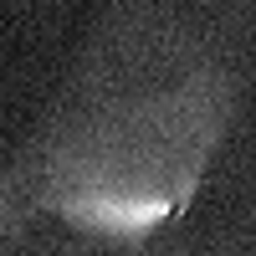}};
				\end{tikzpicture} \hspace{.55cm}
				&
				\begin{tikzpicture}
					\node {\includegraphics[width=0.2\textwidth]{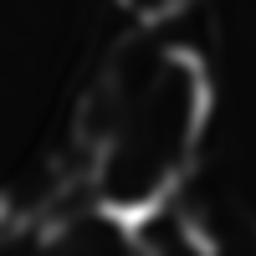}};
				\end{tikzpicture} \hspace{.55cm}
				&       
				\begin{tikzpicture}  
					\node {\includegraphics[width=0.2\textwidth]{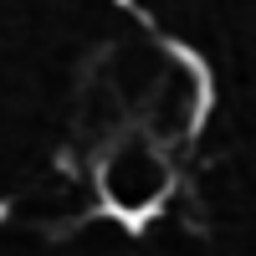}};
				\end{tikzpicture} \hspace{.55cm}
				&       
				\begin{tikzpicture}
					\node {\includegraphics[width=0.2\textwidth]{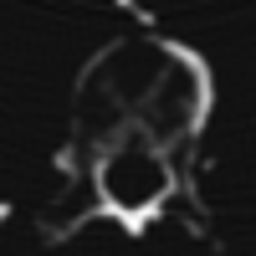}};
				\end{tikzpicture} \hspace{.55cm}
			\end{tabular}
		}
		\caption{Qualitative decomposition results for soft tissue and bone using: (a) FBP, (b) MCG diffusion, (c) DIRECT-Net, (d) proposed \ac{DMDL} method and (e) ground truth.}\label{fig:result_compare}
	\end{center}
\end{figure*}

\subsection{Numerical results}

\noindent The proposed DEcomp-MoD method is compared with: 

\begin{itemize}
    \item FBP: We first applied the material decomposition U-Net to obtain material sinograms and then directly used FBP with Ram-Lak filter to reconstruct material images. Comparison between FBP and proposed DEcomp-MoD can be also regarded as an ablation study of diffusion model and it shows how the diffusion model improves the image quality. 
    \item Manifold Constraint Gradient (MCG) diffusion model: it was proposed in \cite{chung2022improving} to solve inverse problems including CT reconstruction. MCG diffusion model was implemented using the official code and the model trained on CT images from \cite{chung2022improving}. As the model takes image as input, we directly applied FBP on the decomposed sinogram from U-Net and input the reconstructed material image into MCG diffusion model.
    \item DIRECT-Net: it is based on the use of a mutual-domain network to achieve material decomposition in one step \cite{su2022}. We implemented the DIRECT-Net by following the description of network architecture and training details in the article \cite{su2022}. 
\end{itemize}

Fig. \ref{fig:result_compare} shows the results of our proposed method and the comparison. The zoomed region of interest (ROI) is displayed at the bottom of the material image. As we simulated the data acquisition in a sparse-view case, the FBP results of both water and bone materials are corrupted by artifacts. MCG diffusion model removes the streaking artifacts but the results are noisy and the generated structure heavily deviates from the ground truth. DIRECT-Net has visually good denoising performance due to its end-to-end mutual domain network architecture but the results are not accurate as in the sparse-view case, errors are propagated during domain transformation. DEcomp-MoD produces more accurate material decomposition results both qualitatively and quantitatively.

\definecolor{Gray}{gray}{0.9}
\newcolumntype{g}{>{\columncolor{Gray}}c}
\begin{table}[!h]
	\caption{Quantitative results in terms of PSNR and SSIM for soft tissue and bone decomposition.}\label{tab:compare_metrics}
	\begin{center}
		\setlength{\tabcolsep}{8pt}  
		\begin{tabular}{c|cccg}
			\hline
			\multicolumn{5}{c}{Water}  \\ \hline
			& FBP   & MCG & DIRECT-Net & DEcomp-MoD \\ \hline
			PSNR    & $20.50\pm1.21$ & 18.63 & $23.28\pm0.91$ & $\mathbf{29.14\pm 0.78}$     \\
			SSIM    & $0.64\pm0.02$  & 0.51  & $0.79\pm0.02$  & $\mathbf{0.85\pm 0.01}$       
			\\ \hline\hline
			\multicolumn{5}{c}{Bone}  \\ \hline
			& FBP   & MCG & DIRECT-Net & DEcomp-MoD \\ \hline
			PSNR    & $27.26\pm2.10$ & 19.81 & $26.67\pm1.66$ & $\mathbf{33.88\pm3.68}$      \\
			SSIM    & $0.67\pm0.06$  & 0.29  & $0.75\pm0.07$  & $\mathbf{0.87\pm0.06}$      
			\\ \hline
		\end{tabular}
	\end{center}
\end{table}

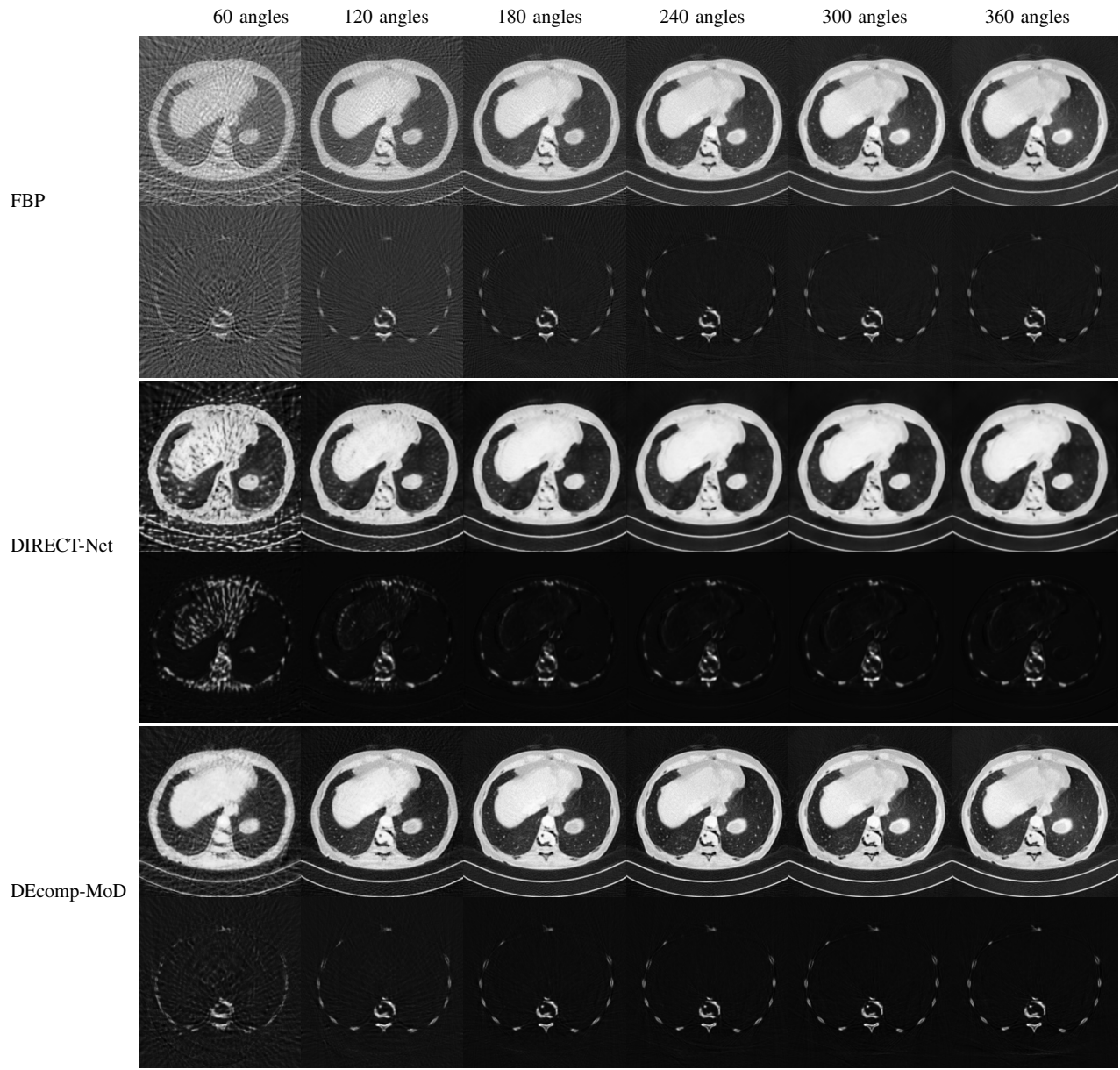
\begin{figure*}[!h]
	\centering
	\small\addtolength{\tabcolsep}{-15pt}
	\resizebox {.95\textwidth} {.95\height} {
		\begin{tabular}{lcccccc}
			& 
			\specialcell{\hspace{.6cm}\small 60 angles}
			&
			\specialcell{\hspace{-.3cm}\small 120 angles}
			& 
			\specialcell{\hspace{-.6cm}\small 180 angles} 
			& 
			\specialcell{\hspace{-.6cm}\small 240 angles}
			& 
			\specialcell{\hspace{-.6cm}\small 300 angles} 
			& 
			\specialcell{\hspace{-.6cm}\small 360 angles} 
			\\
			\vspace{-.32cm}
			\begin{tikzpicture}
				\node[text width = 0.15\textwidth, text depth = 0cm, anchor=north] {\small\hspace{.1cm} FBP};
			\end{tikzpicture} 
			&
			\begin{tikzpicture} 
				\node{\includegraphics[width=0.15\textwidth]{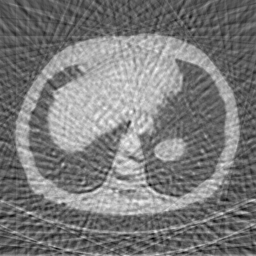}}; 	
			\end{tikzpicture} \hspace{.25cm}
			&
			\begin{tikzpicture} 
				\node{\includegraphics[width=0.15\textwidth]{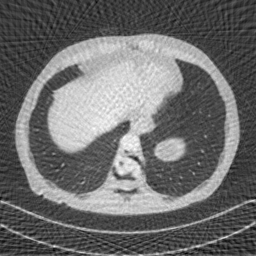}}; 	
			\end{tikzpicture} \hspace{.25cm}
			&
			\begin{tikzpicture} 
				\node{\includegraphics[width=0.15\textwidth]{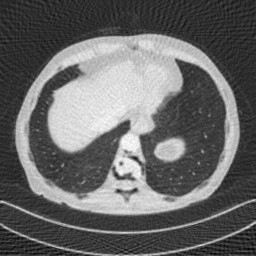}}; 	
			\end{tikzpicture} \hspace{.25cm}
			&
			\begin{tikzpicture} 
				\node{\includegraphics[width=0.15\textwidth]{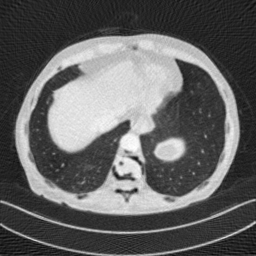}}; 	
			\end{tikzpicture} \hspace{.25cm}
			&
			\begin{tikzpicture} 
				\node{\includegraphics[width=0.15\textwidth]{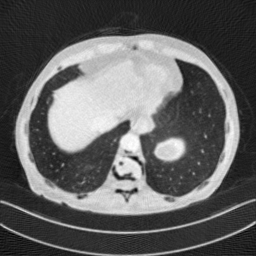}}; 	
			\end{tikzpicture} \hspace{.25cm}
			&
			\begin{tikzpicture} 
				\node{\includegraphics[width=0.15\textwidth]{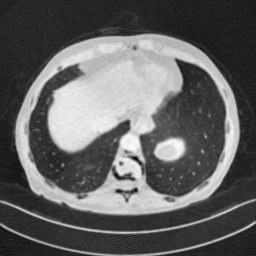}}; 	
			\end{tikzpicture} \hspace{.25cm}
			\\ 
			\vspace{-.25cm}
			\begin{tikzpicture}
				\node[text width = 0.15\textwidth, text depth = 0cm, anchor=north] {};
			\end{tikzpicture} 
			&
			\begin{tikzpicture} 
				\node{\includegraphics[width=0.15\textwidth]{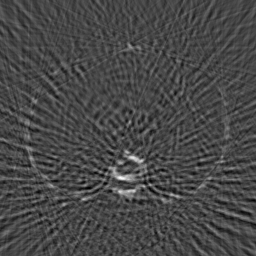}}; 	
			\end{tikzpicture} \hspace{.25cm}
			&
			\begin{tikzpicture} 
				\node{\includegraphics[width=0.15\textwidth]{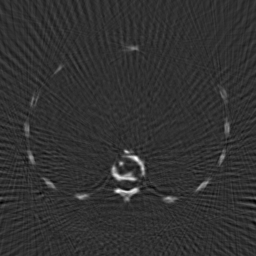}}; 	
			\end{tikzpicture} \hspace{.25cm}
			&
			\begin{tikzpicture} 
				\node{\includegraphics[width=0.15\textwidth]{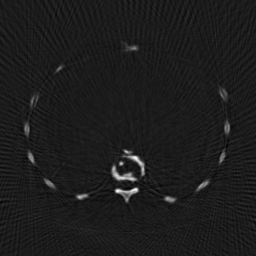}}; 	
			\end{tikzpicture} \hspace{.25cm}
			&
			\begin{tikzpicture} 
				\node{\includegraphics[width=0.15\textwidth]{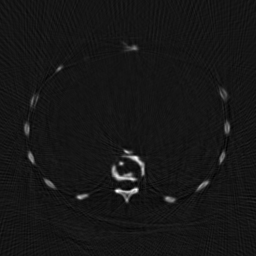}}; 	
			\end{tikzpicture} \hspace{.25cm}
			&
			\begin{tikzpicture} 
				\node{\includegraphics[width=0.15\textwidth]{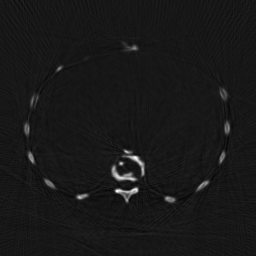}}; 	
			\end{tikzpicture} \hspace{.25cm}
			&
			\begin{tikzpicture} 
				\node{\includegraphics[width=0.15\textwidth]{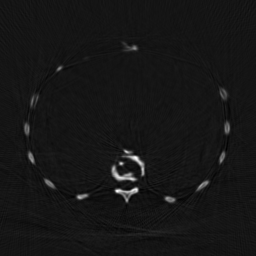}}; 	
			\end{tikzpicture} \hspace{.25cm}
			\\
			\vspace{-.31cm}
			\begin{tikzpicture}
				\node[text width = 0.15\textwidth, text depth = 0cm, anchor=north] {\small\hspace{.1cm} DIRECT-Net};
			\end{tikzpicture} 
			&
			\begin{tikzpicture} 
				\node{\includegraphics[width=0.15\textwidth]{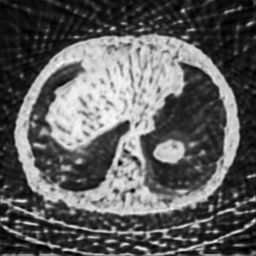}}; 	
			\end{tikzpicture} \hspace{.25cm}
			&
			\begin{tikzpicture} 
				\node{\includegraphics[width=0.15\textwidth]{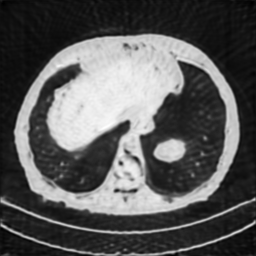}}; 	
			\end{tikzpicture} \hspace{.25cm}
			&
			\begin{tikzpicture} 
				\node{\includegraphics[width=0.15\textwidth]{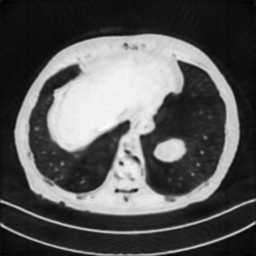}}; 	
			\end{tikzpicture} \hspace{.25cm}
			&
			\begin{tikzpicture} 
				\node{\includegraphics[width=0.15\textwidth]{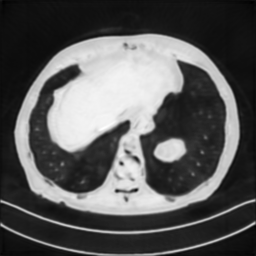}}; 	
			\end{tikzpicture} \hspace{.25cm}
			&
			\begin{tikzpicture} 
				\node{\includegraphics[width=0.15\textwidth]{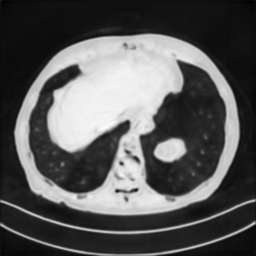}}; 	
			\end{tikzpicture} \hspace{.25cm}
			&
			\begin{tikzpicture} 
				\node{\includegraphics[width=0.15\textwidth]{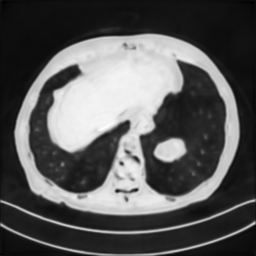}}; 	
			\end{tikzpicture} \hspace{.25cm}
			\\
			\vspace{-.25cm}
			\begin{tikzpicture}
				\node[text width = 0.15\textwidth, text depth = 0cm, anchor=north] {};
			\end{tikzpicture} 
			&
			\begin{tikzpicture} 
				\node{\includegraphics[width=0.15\textwidth]{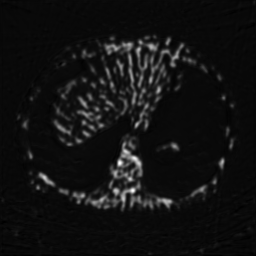}}; 	
			\end{tikzpicture} \hspace{.25cm}
			&
			\begin{tikzpicture} 
				\node{\includegraphics[width=0.15\textwidth]{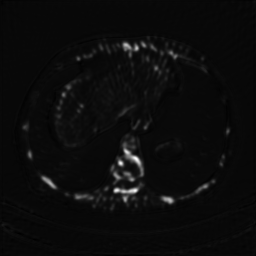}}; 	
			\end{tikzpicture} \hspace{.25cm}
			&
			\begin{tikzpicture} 
				\node{\includegraphics[width=0.15\textwidth]{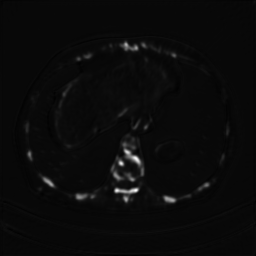}}; 	
			\end{tikzpicture} \hspace{.25cm}
			&
			\begin{tikzpicture} 
				\node{\includegraphics[width=0.15\textwidth]{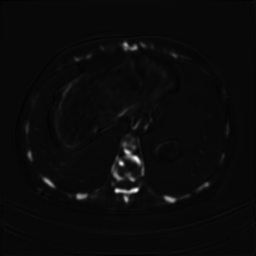}}; 	
			\end{tikzpicture} \hspace{.25cm}
			&
			\begin{tikzpicture} 
				\node{\includegraphics[width=0.15\textwidth]{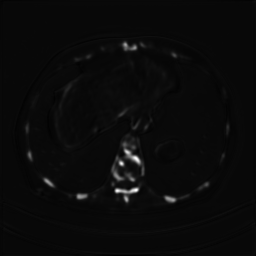}}; 	
			\end{tikzpicture} \hspace{.25cm}
			&
			\begin{tikzpicture} 
				\node{\includegraphics[width=0.15\textwidth]{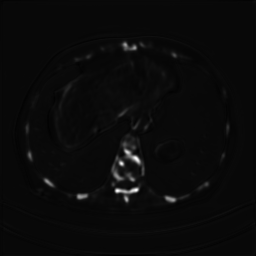}}; 	
			\end{tikzpicture} \hspace{.25cm}
			\\
			\vspace{-.31cm}
			\begin{tikzpicture}
				\node[text width = 0.15\textwidth, text depth = 0cm, anchor=north] {\small\hspace{.1cm} DEcomp-MoD};
			\end{tikzpicture} 
			&
			\begin{tikzpicture} 
				\node{\includegraphics[width=0.15\textwidth]{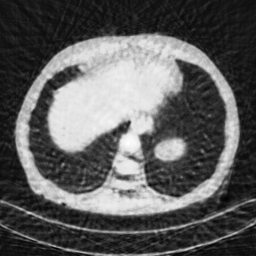}}; 	
			\end{tikzpicture} \hspace{.25cm}
			&
			\begin{tikzpicture} 
				\node{\includegraphics[width=0.15\textwidth]{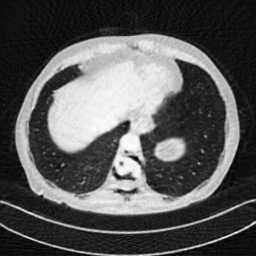}}; 	
			\end{tikzpicture} \hspace{.25cm}
			&
			\begin{tikzpicture} 
				\node{\includegraphics[width=0.15\textwidth]{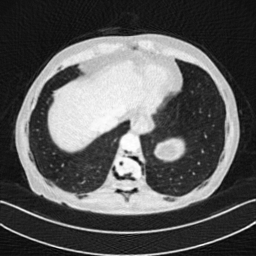}}; 	
			\end{tikzpicture} \hspace{.25cm}
			&
			\begin{tikzpicture} 
				\node{\includegraphics[width=0.15\textwidth]{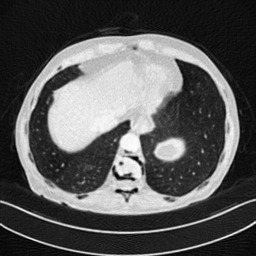}}; 	
			\end{tikzpicture} \hspace{.25cm}
			&
			\begin{tikzpicture} 
				\node{\includegraphics[width=0.15\textwidth]{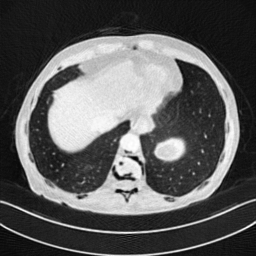}}; 	
			\end{tikzpicture} \hspace{.25cm}
			&
			\begin{tikzpicture} 
				\node{\includegraphics[width=0.15\textwidth]{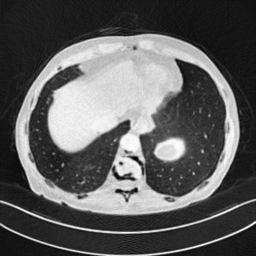}}; 	
			\end{tikzpicture} \hspace{.25cm}
			\\
			\vspace{-.25cm}
			\begin{tikzpicture}
				\node[text width = 0.15\textwidth, text depth = 0cm, anchor=north] {};
			\end{tikzpicture} 
			&
			\begin{tikzpicture} 
				\node{\includegraphics[width=0.15\textwidth]{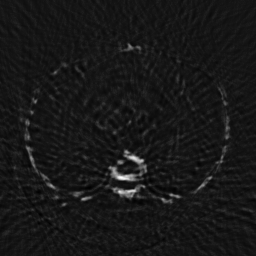}}; 	
			\end{tikzpicture} \hspace{.25cm}
			&
			\begin{tikzpicture} 
				\node{\includegraphics[width=0.15\textwidth]{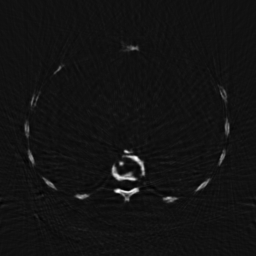}}; 	
			\end{tikzpicture} \hspace{.25cm}
			&
			\begin{tikzpicture} 
				\node{\includegraphics[width=0.15\textwidth]{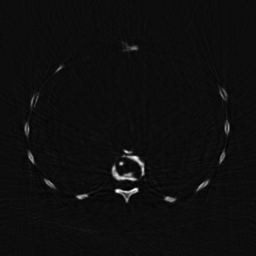}}; 	
			\end{tikzpicture} \hspace{.25cm}
			&
			\begin{tikzpicture} 
				\node{\includegraphics[width=0.15\textwidth]{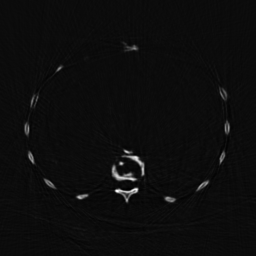}}; 	
			\end{tikzpicture} \hspace{.25cm}
			&
			\begin{tikzpicture} 
				\node{\includegraphics[width=0.15\textwidth]{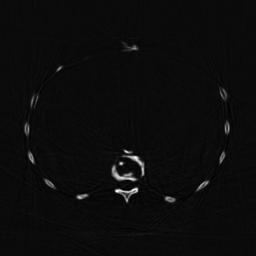}}; 	
			\end{tikzpicture} \hspace{.25cm}
			&
			\begin{tikzpicture} 
				\node{\includegraphics[width=0.15\textwidth]{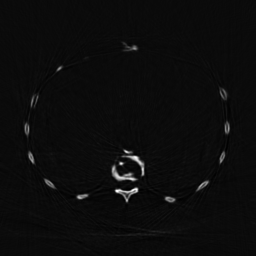}}; 	
			\end{tikzpicture} \hspace{.25cm}
		\end{tabular}
	}
	\caption{Qualitative results of FBP, DIRECT-Net and DEcomp-MoD using measurement projections acquired from different number of angles $[60, 120, 180, 240, 300, 360]$.}
	\label{fig: generalization_images}
\end{figure*}

Table \ref{tab:compare_metrics} presents the PSNR and SSIM metrics of different algorithms. Mean PSNR and SSIM with standard deviation are calculated on the full testing images. Since MCG diffusion model is extremely time-consuming, we only calculate its metrics on the single slice showed in Fig. \ref{fig:result_compare}. Results of both water- and bone-basis material images are presented.

\subsection{Generalization analysis of DEcomp-MoD}\label{sec: generalizaion}

We analysed the generalization performance of FBP, DIRECT-Net and DEcomp-MoD when these algorithms are trained on data obtained from 180 angular projectiond and at inference the CT data is acquired at difference conditions, using fixed photon counts $2\cdot{} 10^6$ and different number of sampling angles $60, 120, 180, 240, 300, 360$. Fig. \ref{fig: generalization_images} shows the qualitative results for bone and water images and it can be clearly noted that DEcomp-MoD has the best generalization performance when compared with other two methods.

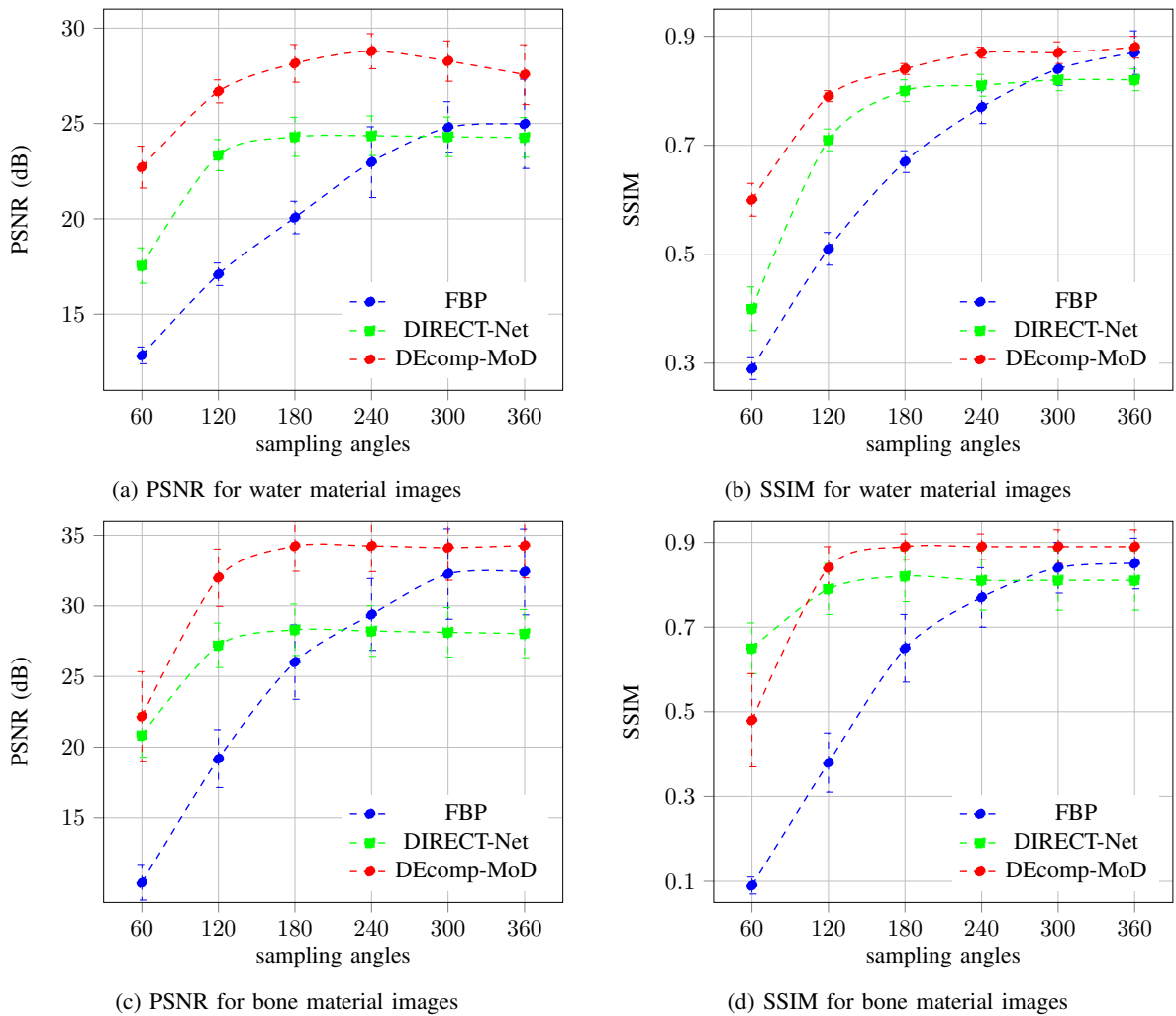
\begin{figure*}[!ht]
	\centering
	\subfloat[PSNR for water material images]{
		\scalebox{.9}{\begin{tikzpicture}
	\begin{axis}[grid, ymin=11, ymax=31, 
		ytick={15, 20.0, 25.0, 30.0}, ytick align=outside, ytick pos=left,
		xtick={60,120,180,240,300,360}, xtick align=outside, xtick pos=left,
		xlabel={sampling angles},
		ylabel={PSNR (dB)},
		legend pos=south east,
		legend style={draw=none}]
		\addplot+[blue, dashed, line width=1.2\pgflinewidth, mark options={blue}, smooth, 
		error bars/.cd, y fixed, y dir=both, y explicit
		] table [x=x, y=y,y error=error, col sep=comma] {
			x,      y,       error
            60,     12.84,   0.44   
            120,    17.09,   0.59    
            180,    20.07,   0.85
            240,    22.97,   1.86
            300,    24.80,   1.34 
            360,    24.98,   2.34 
		};
		\addlegendentry{FBP}
		\addplot+[green, dashed, line width=1.2\pgflinewidth, mark options={green}, smooth, 
		error bars/.cd, y fixed, y dir=both, y explicit
		] table [x=x, y=y,y error=error, col sep=comma] {
			x,      y,           error
			60,     17.55,       0.93   
			120,    23.34,       0.82   
			180,    24.30,       1.03
			240,    24.36,       1.04
			300,    24.30,       1.04
			360,    24.27,       1.04 
		};
		\addlegendentry{DIRECT-Net}
		\addplot+[red, dashed, line width=1.2\pgflinewidth, mark options={red}, smooth, 
		error bars/.cd, y fixed, y dir=both, y explicit
		] table [x=x, y=y,y error=error, col sep=comma] {
			x,      y,           error
			60,     22.71,       1.10  
			120,    26.68,       0.60  
			180,    28.15,       0.99
			240,    28.79,       0.92
			300,    28.27,       1.06
			360,    27.56,       1.57 
		};
		\addlegendentry{DEcomp-MoD}
	\end{axis}
\end{tikzpicture}
	}}
	\quad
	\subfloat[SSIM for water material images]{
		\scalebox{.9}{\begin{tikzpicture}
	\begin{axis}[grid, ymin=0.25, ymax=0.95, 
		ytick={0.3, 0.5, 0.7, 0.9}, ytick align=outside, ytick pos=left,
		xtick={60,120,180,240,300,360}, xtick align=outside, xtick pos=left,
		xlabel={sampling angles},
		ylabel={SSIM},
		legend pos=south east,
		legend style={draw=none}]
		\addplot+[blue, dashed, line width=1.2\pgflinewidth, mark options={blue}, smooth, 
		error bars/.cd, y fixed, y dir=both, y explicit
		] table [x=x, y=y,y error=error, col sep=comma] {
			x,      y,    error
			60,     0.29, 0.02   
			120,    0.51, 0.03    
			180,    0.67, 0.02
			240,    0.77, 0.03
			300,    0.84, 0.03 
			360,    0.87, 0.04 
		};
		\addlegendentry{FBP}
		\addplot+[green, dashed, line width=1.2\pgflinewidth, mark options={green}, smooth, 
		error bars/.cd, y fixed, y dir=both, y explicit
		] table [x=x, y=y,y error=error, col sep=comma] {
			x,      y,           error
			60,     0.40,        0.04   
			120,    0.71,        0.02   
			180,    0.80,        0.02
			240,    0.81,        0.02
			300,    0.82,        0.02
			360,    0.82,        0.02 
		};
		\addlegendentry{DIRECT-Net}
		\addplot+[red, dashed, line width=1.2\pgflinewidth, mark options={red}, smooth, 
		error bars/.cd, y fixed, y dir=both, y explicit
		] table [x=x, y=y,y error=error, col sep=comma] {
			x,      y,           error
			60,     0.60,        0.03 
			120,    0.79,        0.01  
			180,    0.84,        0.01
			240,    0.87,        0.01
			300,    0.87,        0.02
			360,    0.88,        0.02 
		};
		\addlegendentry{DEcomp-MoD}
	\end{axis}
\end{tikzpicture} 
	}} \\
	\vspace{.2cm}
	\subfloat[PSNR for bone material images]{
		\scalebox{.9}{\begin{tikzpicture}
	\begin{axis}[grid, ymin=9, ymax=36, 
		ytick={15, 20.0, 25.0, 30.0, 35.0}, ytick align=outside, ytick pos=left,
		xtick={60,120,180,240,300,360}, xtick align=outside, xtick pos=left,
		xlabel={sampling angles},
		ylabel={PSNR (dB)},
		legend pos=south east,
		legend style={draw=none}]
		\addplot+[blue, dashed, line width=1.2\pgflinewidth, mark options={blue}, smooth, 
		error bars/.cd, y fixed, y dir=both, y explicit
		] table [x=x, y=y,y error=error, col sep=comma] {
			x,      y,     error
            60,     10.40, 1.23 
            120,    19.18, 2.05    
            180,    26.02, 2.64
            240,    29.39, 2.54
            300,    32.26, 3.20 
            360,    32.41, 3.04 
		};
		\addlegendentry{FBP}
		\addplot+[green, dashed, line width=1.2\pgflinewidth, mark options={green}, smooth, 
		error bars/.cd, y fixed, y dir=both, y explicit
		] table [x=x, y=y,y error=error, col sep=comma] {
			x,      y,           error
			60,     20.84,       1.56  
			120,    27.21,       1.58   
			180,    28.31,       1.82
			240,    28.22,       1.79
			300,    28.13,       1.75
			360,    28.04,       1.71
		};
		\addlegendentry{DIRECT-Net}
		\addplot+[red, dashed, line width=1.2\pgflinewidth, mark options={red}, smooth, 
		error bars/.cd, y fixed, y dir=both, y explicit
		] table [x=x, y=y,y error=error, col sep=comma] {
			x,      y,           error
			60,     22.17,       3.17 
			120,    32.00,       2.03 
			180,    34.24,       1.80
			240,    34.25,       1.83
			300,    34.13,       2.32
			360,    34.28,       2.30
		};
		\addlegendentry{DEcomp-MoD}
	\end{axis}
\end{tikzpicture}
	}}
	\quad
	\subfloat[SSIM for bone material images]{
		\scalebox{.9}{\begin{tikzpicture}
	\begin{axis}[grid, ymin=0.05, ymax=0.95, 
		ytick={0.1, 0.3, 0.5, 0.7, 0.9}, ytick align=outside, ytick pos=left,
		xtick={60, 120, 180, 240, 300, 360}, xtick align=outside, xtick pos=left,
		xlabel={sampling angles},
		ylabel={SSIM},
		legend pos=south east,
		legend style={draw=none}]
		\addplot+[blue, dashed, line width=1.2\pgflinewidth, mark options={blue}, smooth, 
		error bars/.cd, y fixed, y dir=both, y explicit
		] table [x=x, y=y,y error=error, col sep=comma] {
			x,      y,    error
			60,     0.09, 0.02   
			120,    0.38, 0.07    
			180,    0.65, 0.08
			240,    0.77, 0.07
			300,    0.84, 0.06 
			360,    0.85, 0.06 
		};
		\addlegendentry{FBP}
		\addplot+[green, dashed, line width=1.2\pgflinewidth, mark options={green}, smooth, 
		error bars/.cd, y fixed, y dir=both, y explicit
		] table [x=x, y=y,y error=error, col sep=comma] {
			x,      y,           error
			60,     0.65,        0.06   
			120,    0.79,        0.06   
			180,    0.82,        0.06
			240,    0.81,        0.07
			300,    0.81,        0.07
			360,    0.81,        0.07 
		};
		\addlegendentry{DIRECT-Net}
		\addplot+[red, dashed, line width=1.2\pgflinewidth, mark options={red}, smooth, 
		error bars/.cd, y fixed, y dir=both, y explicit
		] table [x=x, y=y,y error=error, col sep=comma] {
			x,      y,           error
			60,     0.48,        0.11 
			120,    0.84,        0.05  
			180,    0.89,        0.03
			240,    0.89,        0.03
			300,    0.89,        0.04
			360,    0.89,        0.04 
		};
		\addlegendentry{DEcomp-MoD}
	\end{axis}
\end{tikzpicture} 
	}}
	\caption{Generalization of FBP, DIRECT-Net and DEcomp-MoD. The three algorithms are trained with data acquired by $2\cdot{} 10^6$ photons and $180$ sampling angles. The PSNR and SSIM are calculated on testing data acquired from different number of sampling angles.} \label{fig: generalization_plot}
\end{figure*}

When the data are sampled by only 60 angles, FBP and DIRECT-Net produce extremely noisy results and the material decomposition of DIRECT-Net is not accurate. With other sampling angles, DEcomp-MoD has more stable results and higher metrics than other two methods. DIRECT-Net is trained in a traditional supervised manner that a pair of low- and high- dose CT images are required and this kind of supervised method usually has poor generalization ability when the testing conditions are different. While the training of diffusion model does not need paired data and diffusion model has much better generalization ability. Though DEcomp-MoD uses supervised U-Net to decompose material in projection domain, with the pre-trained diffusion model and known material sinograms, DEcomp-MoD still can suppress the noise in a wide range of sampling angles. 

The quantitative results in terms of PSNR and SSIM are shown in Fig. \ref{fig: generalization_plot} for water (top) and bone (bottom) for different sampling angles and it highlights how DEcomp-MoD has superior performance, consistently over several set of angles, simulating different amount of X-ray dose, on average $5$ dB higher PSNR compared to DIRECT-Net and $9$ dB over FBP.

\subsection{Analysis of DEcomp-MoD Parameters}\label{sec: parameters}

\subsubsection{Effect of $\lambda$ and $\xi$} \label{sec: free_parameters}

\begin{figure*}[!h]
	\begin{center}	
		\includegraphics[width=0.8\textwidth]{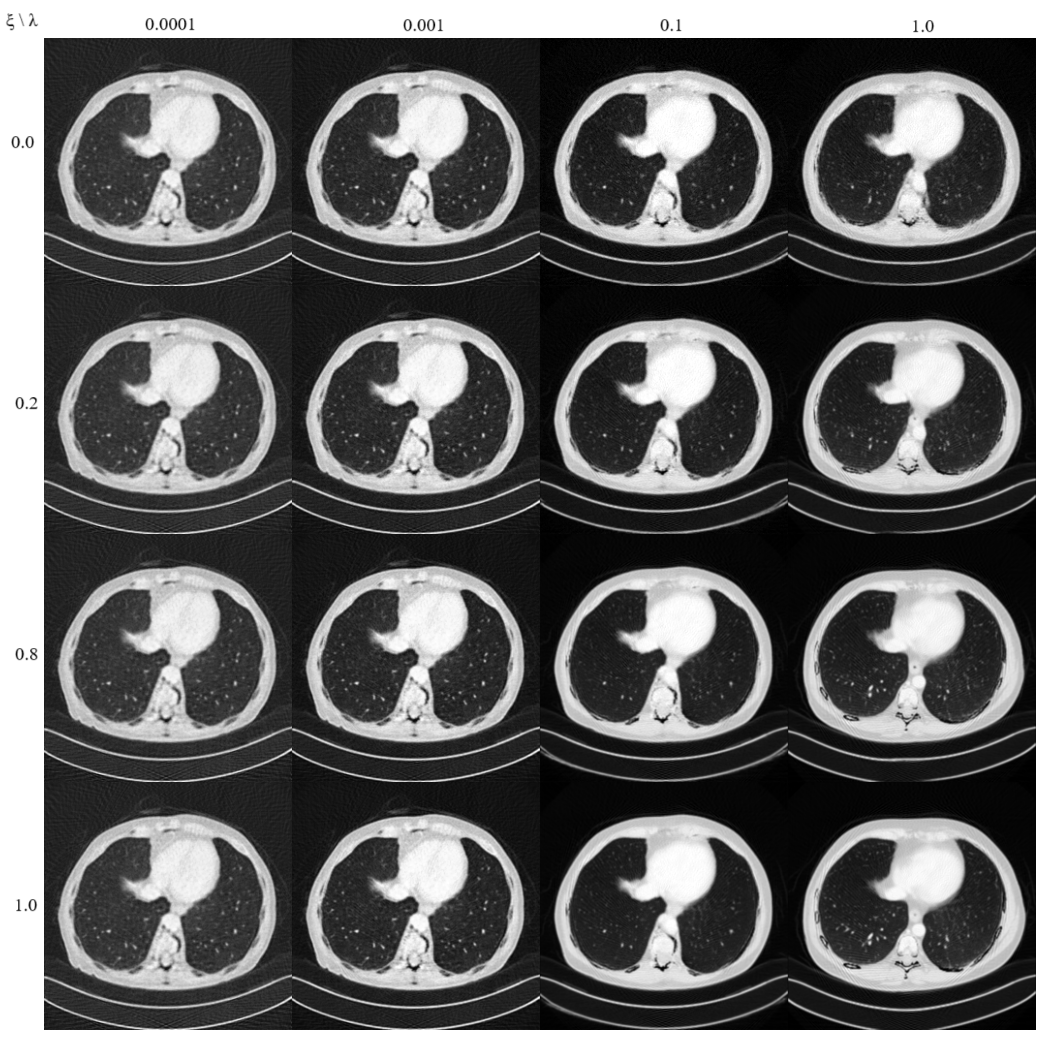}
		\caption{Qualitative results of DEcomp-MoD for different set of parameters $\lambda = [10^{-4}, 10^{-3}, 0.1, 1]$ and $\xi = [0, 0.2, 0.8, 1]$.}
		\label{fig:parameters}
	\end{center}
\end{figure*}

We analysed the effect of the choice of the parameters $\lambda$ and $\xi$ of DEcomp-MoD algorithm on the testing performance. Based on Algorithm \ref{table:DMDL}, the parameter $\lambda$ controls the strength of diffusion model and $\xi$ controls the weight of current noise and random noise during reverse sampling process \cite{Zhu2023}. Fig. \ref{fig:parameters} shows the qualitative results on a reconstructed CT material image with different values of $\lambda$ (across columns) and $\xi$ (across rows).  It is possible to note that as $\lambda$ is increasing, the strength of diffusion model is higher and the image retains less blurred effects, but the anatomical structure deviates more from the ground truth. When $\xi =0$, the image generation is determined only by the effective noise at current time-step $t$. Therefore, the anatomical results are closed to the ground truth but images are noisy. When $\xi =1$, the image generation is determined only from the normal distribution, and the results are cleaner and smooth but possibly deviate from ground truth if $\lambda$ is inappropriate. Based on the qualitative and quantitative results, we selected $\lambda =0.001$ and $\xi =1.0$ as default baseline. 

\subsubsection{Effect of sampling iterations} \label{sec: free_iters}

We considered the number of iterations of the reverse diffusion process at testing and we evaluated the trade-off between computational time and accuracy. Table \ref{tab:compare_iters} shows the quantitative results in terms of accuracy, measured by the PSNR, and running time for the reconstruction of one CT material image. Using on a PC with 16 CPU cores, DEcomp-MoD required $9.42$ seconds to generate one CT image after $100$ sampling steps. Although FBP is 40\% faster in computational time, the achievable accuracy with sparse-views is not clinically acceptable, as shown in Fig. \ref{fig: generalization_images}. Furthermore, the computational cost of DEcomp-MoD is comparable with unrolling deep learning algorithms but much faster compared to DDPM diffusion algorithm. 

\begin{table}[!h]
	\caption{Quantitative results in terms of PSNR, SSIM and running time for generating a single slice of material image with different sampling iterations.}\label{tab:compare_iters}
	\begin{minipage}{.5\textwidth}
        \begin{center}
     	\setlength{\tabcolsep}{8pt}  
		  \begin{tabular}{ccgcc}
            \hline
            
            \multicolumn{1}{c|}{iterations}     &  10  & 100 & 500 & 1000 \\ \hline
            \multicolumn{1}{c|}{PSNR}    & 20.76 & $\mathbf{29.48}$ & 28.31 & 27.86     \\
            \multicolumn{1}{c|}{SSIM}    & 0.72  & $\mathbf{0.85}$ & 0.85 & 0.84       \\
            \multicolumn{1}{c|}{Time}    & \textbf{2.74 sec}  & 9.42 sec  & 43.28 sec  & 82.42 sec \\ \hline
            \end{tabular}
        \end{center}
	\end{minipage}
\end{table}

\begin{figure*}[!h]
	\begin{center}
		\small\addtolength{\tabcolsep}{-18pt}
		\renewcommand{\arraystretch}{0.1}
		\resizebox {.95\textwidth} {.95\height}{
			\begin{tabular}{ccccc}
				\hspace{-.8cm}\footnotesize{(a) $T=10$} 
				& 
				\hspace{-.8cm}\footnotesize (b) $T=100$
				& 
				\hspace{-.8cm}\footnotesize (c) $T=500$
				& 
				\hspace{-.8cm}\footnotesize (d) $T=1000$
				& 
				\hspace{-.8cm}\footnotesize (e) Ground truth \\
				\vspace{.2cm} 
				\begin{tikzpicture}  
					\node {\includegraphics[width=0.2\textwidth]{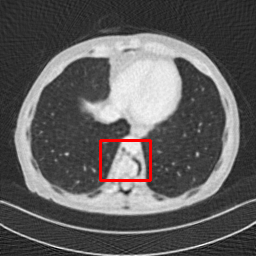}};
				\end{tikzpicture} \hspace{.55cm}
				&
				\begin{tikzpicture}
					\node {\includegraphics[ width=0.2\textwidth]{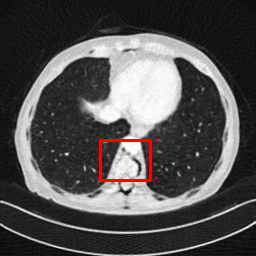}};
				\end{tikzpicture} \hspace{.55cm}
				&
				\begin{tikzpicture}
					\node {\includegraphics[width=0.2\textwidth]{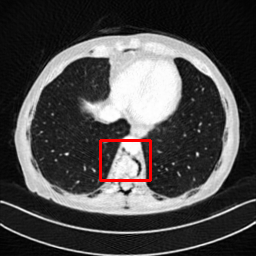}};
				\end{tikzpicture} \hspace{.55cm}
				&       
				\begin{tikzpicture}  
					\node {\includegraphics[width=0.2\textwidth]{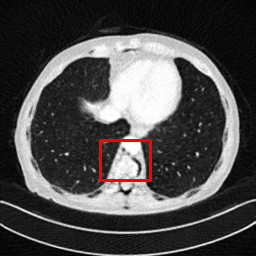}};
				\end{tikzpicture} \hspace{.55cm}
				&       
				\begin{tikzpicture}
					\node {\includegraphics[width=0.2\textwidth]{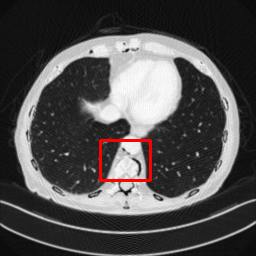}};
				\end{tikzpicture} \hspace{.55cm} \vspace{-.4cm}
				\\   	 
				\begin{tikzpicture}
					\node {\includegraphics[width=0.2\textwidth]{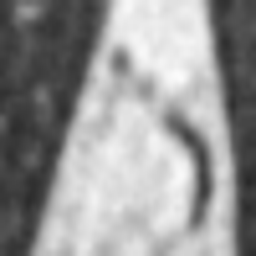}};
				\end{tikzpicture} \hspace{.5cm}
				& 
				\begin{tikzpicture}  
					\node {\includegraphics[width=0.2\textwidth]{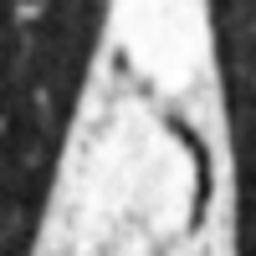}};
				\end{tikzpicture} \hspace{.55cm}
				&
				\begin{tikzpicture}
					\node {\includegraphics[ width=0.2\textwidth]{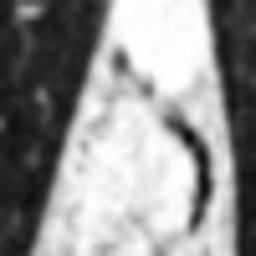}};
				\end{tikzpicture} \hspace{.55cm}
				&      
				\begin{tikzpicture}  
					\node {\includegraphics[width=0.2\textwidth]{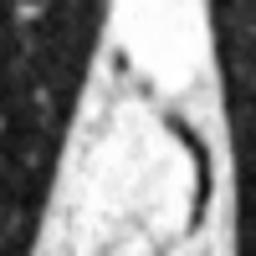}};
				\end{tikzpicture} \hspace{.55cm}
				&       
				\begin{tikzpicture}
					\node {\includegraphics[width=0.2\textwidth]{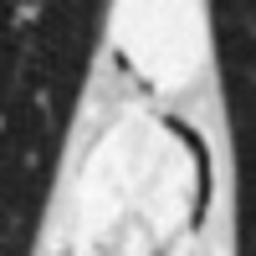}};
				\end{tikzpicture} \hspace{.65cm}
			\end{tabular}
		}
		\caption{Reverse inference process of DEcomp-MoD using different number of sampling iterations $T$.}\label{fig:result_iterations}
	\end{center}
\end{figure*}

Fig. \ref{fig:result_iterations} shows the results at $T = 10, 100, 500, 1000$ iterations. It is worth noting that after $T=10$ iterations the global anatomical structure of the image is correctly captured but the details are blurred with several artefacts. However, already after $T=100$ iterations of DEcomp-MoD the image retains all local details with an accuracy which is visually comparable to the ones obtained after $T=500, 1000$ iterations.

\section{Discussions} \label{sec: discussion}

\subsubsection{DEcomp-MoD Parameters Settings}

In Algorithm \ref{table:DMDL}, $\mu_t = \lambda/\bar{\sigma}_t^2$ is the parameter that controls the weight of the prior term which is learned from the reverse diffusion process. $\lambda$ is a free parameter and is set to $0.001$ according to the analysis in section \ref{sec: free_parameters}. The vector of parameters $\bar{\sigma}_t^2$ is pre-calculated as it represent the known variance schedule. 

Therefore, as $\lambda$ increases the weight of the diffusion image generation in DEcomp-MoD increases, and vice versa. We have shown in section \ref{sec: free_parameters} the effect of $\lambda$. When $\lambda$ is small, the generative power of the diffusion model is weak and the artifacts in the CT images are not removed effectively. In contrast, the diffusion model generates other tissues when $\lambda$ is too large. In our case, values around $\lambda =0.001$ have empirically shown the best result. 

$\xi$ is the parameter controlling the weight of the effective noise and random noise in the reverse sampling of the diffusion model. The result becomes smoother as $\xi$ increases and we set the default value $\xi =1.0$ based on the results in section \ref{sec: free_parameters}.

\subsubsection{DEcomp-MoD Superior Accuracy}

Previous results have shown that our proposed DEcomp-MoD method can produce high-quality material decomposition results. We implemented a projection-based material decomposition in combination with the powerful generative and denoising ability of denoising diffusion model, that allows to recover a material images with higher qualitative and quantitative accuracy compared to one-step methods and alternative diffusion models. 

In addition, the sampling time of our proposed diffusion model using CG is comparable to iterative reconstruction and it is faster compared to other diffusion models (MCG) for CT reconstruction. DEcomp-MoD only needs $10$ seconds to generate a 256\texttimes{} 256 CT image while MCG diffusion model needs few minutes.

\subsubsection{DEcomp-MoD Future Improvements}

There are potential improvement to the current DEcomp-MoD version that can be developed in the future. To begin with, we currently have to train two models separately which is inefficient. 

Moreover, currently we can only generate one material at a time, so we have to repeat the process multiple times for multiple materials. The limitations can be solved by developing a multi-channel structure and combining two training phases into an end-to-end training scheme.

In our supervised training strategy, we do not need pair of images to train the diffusion model; this means that the pre-trained diffusion model can generalize to images obtained from different CT scanner geometries and X-ray doses. This approach can be extended to unsupervised training exploiting ideas from one-step material decomposition as in \cite{fang2021}, which required much larger computation than our projection-based decomposition using a simple U-Net. However, the unsupervised decomposition may increase the generalization ability when incorporated with the diffusion model which can be considered as a direction of future work.


\section{Conclusion}

We proposed a new material decomposition method DEcomp-MoD for DECT by incorporating the knowledge of the spectral model DECT system into a supervised deep learning training strategy for projection-based material decomposition and combining an unsupervised score-based denoising diffusion learned prior in the material image domain for image reconstruction and restoration. DEcomp-MoD improves both quantitative the qualitative results compared with the existing diffusion-based and supervised methods.

\section*{Acknowledgment}
All authors declare that they have no known conflicts of interest in terms of competing financial interests or personal relationships that could have an influence or are relevant to the work reported in this paper. 
A. Perelli acknowledges the support of the Royal Academy of Engineering under the RAEng / Leverhulme Trust Research Fellowships programme (award LTRF-2324-20-160). The AAPM low-dose CT dataset obtained ethical approval by Mayo Clinic (USA) and it is made available through TCIA restricted license with de-identifiability of the data which is agreed by authors.

\bibliographystyle{IEEEtran}
\bibliography{biblio}

\end{document}